\documentclass[sigconf]{acmart}
\AtBeginDocument{%
  }

\copyrightyear{2025}
\acmYear{2025}
\setcopyright{cc}
\setcctype{by}
\acmConference[CI 2025]{Collective Intelligence Conference}{August 4--6, 2025}{San Diego, CA, USA}
\acmBooktitle{Collective Intelligence Conference (CI 2025), August 4--6, 2025, San Diego, CA, USA}
\acmDOI{10.1145/3715928.3737481}
\acmISBN{979-8-4007-1489-4/2025/08}

\usepackage{longtable}
\usepackage{color}
\usepackage{longtable}
\usepackage{amsmath}
\usepackage{subcaption}
\usepackage{natbib}
\usepackage{booktabs}
\usepackage{float}
\usepackage{hyperref}

\newcommand{\MakePrompt}[3][1]{%
    \begin{center}
        \begin{tabular}{p{#1\linewidth}}
            \texttt{#2} \texttt{#3} \\
        \end{tabular}
    \end{center}
}

\usepackage{color}
\usepackage{framed}

\begin{document}

\title[How AI Ideas Affect Human Ideas]{How AI Ideas Affect the Creativity, Diversity, and Evolution of Human Ideas: Evidence From a Large, Dynamic Experiment}

\author{Joshua Ashkinaze}
\affiliation{%
  \institution{University of Michigan}
  \country{United States}}
\email{jashkina@umich.edu}

\author{Julia Mendelsohn}
\affiliation{%
  \institution{University of Maryland}
  \country{United States}}
\email{juliame@umd.edu}

\author{Li Qiwei}
\affiliation{%
  \institution{University of Michigan}
  \country{United States}}
\email{rrll@umich.edu}

\author{Ceren Budak}
\affiliation{%
  \institution{University of Michigan}
  \country{United States}}
\email{cbudak@umich.edu}

\author{Eric Gilbert}
\affiliation{%
  \institution{University of Michigan}
  \country{United States}}
\email{eegg@umich.edu}

\renewcommand{\shortauthors}{Ashkinaze et al.}

\begin{abstract}

Exposure to large language model output is rapidly increasing. How will seeing AI-generated ideas affect human ideas? We conducted a dynamic experiment (800+ participants, 40+ countries) where participants viewed creative ideas that were from ChatGPT or prior experimental participants, and then brainstormed their own idea. We varied the number of AI-generated examples (none, low, or high \textit{exposure}) and if the examples were labeled as ``AI''  (\textit{disclosure}). We find that high AI exposure (but not low AI exposure) did not affect the creativity of individual ideas but did increase the average amount and rate of change of collective idea diversity. AI made ideas different, not better. There were no main effects of disclosure. We also found that self-reported creative people were less influenced by knowing an idea was from AI and that participants may knowingly adopt AI ideas when the task is difficult. Our findings suggest that introducing AI ideas may increase collective diversity but not individual creativity. 

\end{abstract}



\maketitle

\section{Introduction}

If we think of culture as a feedback loop where individuals and societies shape each other through exchanges of ideas and practices \citep{richerson_not_2008,boyd_culture_1988}, then what happens when generative AI enters this ``culture loop''? Exposure to LLMs (large language models) is increasing: When released, ChatGPT was the fastest-growing consumer application in history \citep{hu_chatgpt_2023}. Moreover, we are likely exposed to even more AI content than we realize: Humans overestimate their ability to distinguish AI from human content \citep{jakesch_human_2022}. This exposure likely matters: Ideas we see affect the ideas we create \citep{nijstad_how_2006}. How, then, will the rapid rise of exposure to LLM-generated ideas affect the creativity, diversity, and evolution of human ideas? And to what extent \textit{do} AI ideas influence human ideas? These questions have implications for how AI may shape collective intelligence~\cite{peeters_hybrid_2021, cui_ai-enhanced_2024, burton_how_2024}. Our study speaks to these questions in the specific context of brainstorming, an activity that occurs across many domains. 

The widespread dissemination of AI-generated content creates a phenomenon we call ``passive exposure,'' which differs from the ``active engagement'' primarily studied in prior human-AI interactions. By ``passive exposure'', we refer to cases when (A) users see LLM outputs but do not have an active role in the creation of these outputs and (B) users are given no instructions to actively engage with these outputs. Passive exposure approximates how users often encounter LLM outputs in the real world. For example, OpenAI users generate 100 billion words per day \cite{griffin_chatgpt_2024}. It is likely that the number of people who are merely seeing (i.e., \textit{passively exposed}) AI output is significantly larger than the number of people who are creating (i.e., \textit{actively engaging}) with these systems. Yet in existing studies of human-AI creativity, participants are often \textit{actively} interacting with an AI system \citep{yang_ai_2022, osone_buncho_2021, lee_coauthor_2022, branch_collaborative_2021, gero_metaphoria_2019, padmakumar_does_2024}. 

As AI exposure has increased, so have concerns over AI \textit{disclosure} \citep{hancock_ai-mediated_2020} (whether providers should disclose when they use AI systems). California considered requiring disclosure on behalf of anyone using bots on social media \citep{williams_should_2018}. Concerns regarding the disclosure of LLMs are only more likely to grow. LLM output is increasingly indistinguishable~\citep{jakesch_human_2022} from that of humans. We are interested if disclosing ideas as coming from AI moderates the effect of AI exposure. 

\begin{figure*}[h]
    \centering
    \includegraphics[width=1\textwidth]{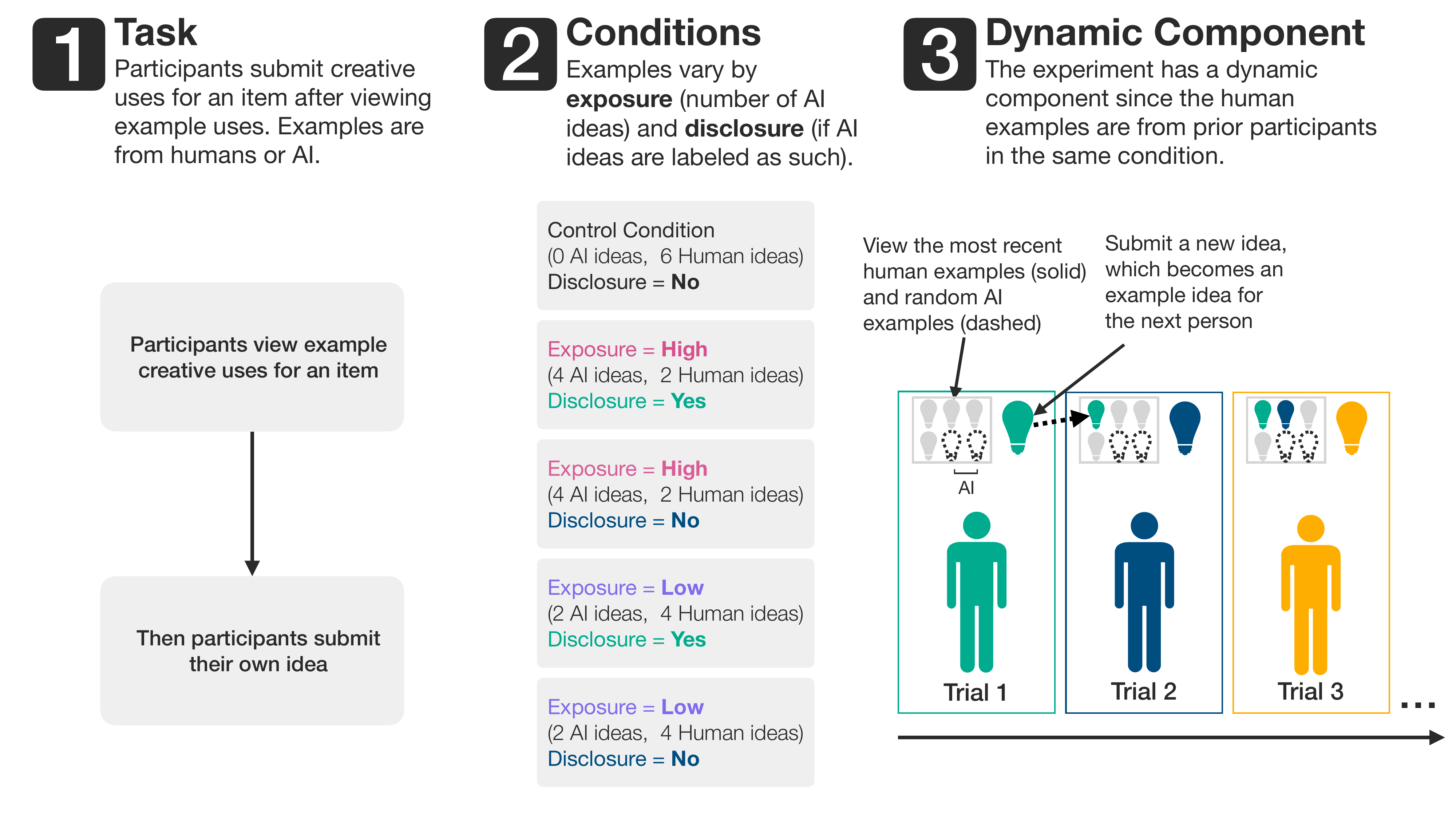}
    \caption{Graphical depiction of experiment. The task (Panel 1) is to submit a creative idea after seeing examples, where examples are from humans or AI. We vary (Panel 2) the amount of AI ideas in the example set (\textit{exposure}) and if AI ideas are labeled as such (\textit{disclosure}). The experiment is dynamic (Panel 3). Responses from prior participants serve as examples for future participants. }
    \label{fig:clean_screenshot}
\end{figure*}

Motivated by these dynamics, we conducted a large-scale experiment to systematically examine how AI exposure and disclosure influence the creativity, diversity, and evolution of human ideas. We utilize a variant of the Alternate Uses Task (AUT, \cite{guilford_alternate_1978}), a standard measure of creativity, and manipulate exposure to LLM-generated ideas. In the AUT, participants generate non-obvious uses for a common item. For example: \textit{What is a creative use for a tire?} In our variant, participants complete the AUT after viewing example ideas, which serve as our experimental manipulation. Examples vary in AI exposure (none, few, or many AI examples) and AI disclosure (whether AI-generated ideas are explicitly labeled). The human-generated ideas in each example set originate from prior participants in the same experimental condition. See Figure \ref{fig:clean_screenshot} for a depiction. 

Our dynamic experiment design leverages ideas from prior participants as stimuli for future participants, reflecting the interdependent nature of brainstorming. Ideas build upon previous ideas. This approach captures (certain aspects of) the compounding effects of integrating LLMs into the ``culture loop'' and simulates potential futures for human-AI collaboration. Our design enables us to observe both average performance levels and temporal dynamics of creativity and diversity across conditions. This setup provides insights into how LLMs shape collective thought.

\vspace{-0.7em}

\subsection{Findings}

Our experiment manipulates two key factors: (1) AI exposure (the number of AI-generated ideas in example sets) and (2) AI disclosure (whether AI ideas are labeled as such). We investigate three outcomes: individual creativity, idea diversity (measured at individual and collective levels), and AI idea adoption (who and when adopts AI example ideas). For creativity and diversity, we analyze both average levels and temporal evolution across experimental iterations. See \autoref{big_table} for our measures. Here are our findings.

\begin{enumerate}

    \item \textbf{High AI exposure increases collective diversity but not individual creativity.} We find that AI exposure did not affect individual creativity. However, conditions with high levels of AI exposure had more \textit{collective} idea diversity. That is, ideas in the high AI exposure conditions were more \textit{different} from each other but not necessarily better. Our findings around creativity and diversity suggest the effect of AI exposure may be nuanced. 
    
    \item \textbf{High AI exposure increases the speed at which idea diversity develops.}  Through our dynamic design, we find that high AI exposure increases not only the average levels of collective idea diversity but also the rate of change in idea diversity. This is a consequential finding since even small differences in rates of change can lead to large cumulative differences over time. 
    
    \item \textbf{People who identify as creative are less influenced by AI disclosure}. We find that for users who self-identify as highly creative, adoption of AI ideas is not influenced by AI disclosure. But AI disclosure did affect the adoption of AI ideas for users who self-identified as low in creativity. This suggests that highly creative people will not be ``duped'' into adopting AI ideas. 

    \item \textbf{Participants may adopt AI ideas for harder prompts.} We find that when AI ideas are disclosed as such, participants were more likely to adopt the ideas of AI for difficult AUT prompts. This suggests that users may rely on AI ideas not for trivial creative tasks but for difficult ones. Since this finding is based on a small number of AUT items, this is speculative.

\end{enumerate}

\subsection{Defining Concepts and Variables} \label{outcome_variables}
\subsubsection{Creativity}
\label{creativity}

Creativity is defined in many ways \citep{walia_dynamic_2019}. But one common conception is divergent thinking \citep{guilford_nature_1967}: When ``an individual solves a problem or reaches a decision using strategies that deviate from commonly used or previously taught strategies'' \citep{american_psychological_association_dictionary_of_psychology_divergent_nodate}. One of the most common \citep{abraham_gender_2016} tests of divergent thinking is the Alternate Uses Task (AUT) \citep{guilford_alternate_1978}\footnote{https://www.mindgarden.com/67-alternate-uses}, where participants are asked to think of an original use for an everyday object. Traditionally, responses to the AUT are measured along four dimensions: originality (how original the idea is), elaboration (how much the participant elaborates on the idea), fluency (how many ideas), and flexibility (different categories of ideas). The latter two can only be measured if the participant provides multiple responses to the same question. Due to our research design\footnote{Participants see the most recent responses in the condition as stimuli, so if one participant brainstorms many responses, that participant would be over-represented in future participants' example sets.}, we have participants generate just one creative idea (as in \citep{beaty_semantic_2022}), and we focus on originality. 

We follow a tradition of scoring responses to the AUT computationally \citep{yu_mad_2023, beaty_automating_2021, beaty_semantic_2022, yang_automatic_2023, organisciak_beyond_2023, dumas_four_2021}. Specifically, we measure the \textbf{creativity} of AUT ideas with an existing fine-tuned GPT-3 classifier \citep{organisciak_beyond_2022}, which has an r=0.81 overall correlation with human judgments of AUT originality. Moreover, we chose AUT items for our experiment where the classifier had the \textit{highest} accuracy\footnote{\textit{tire} (r=0.91), \textit{pants} (r=0.91), \textit{shoe} (r=0.91), \textit{table} (r=0.9), and \textit{bottle} (r=0.88)}. Our task is highly `in-domain' for the classifier: we ask participants to do the \textit{same exact task} for the \textit{same exact items} the model was trained on. We refer to the originality score from this classifier as individual-level \textbf{creativity}, but future work can explore other dimensions of creativity (such as fluency). We discuss this classifier more in Appendix \ref{picking_aut_items}.

\subsubsection{Idea Diversity \& AI Adoption}

In addition to creativity, we measure how our experimental factors (LLM exposure and LLM disclosure) shape the \textit{diversity} of ideas that participants produce. This is a complementary measure to creativity. Creativity is often thought of as an individual outcome. Diversity is a collective outcome. Equivalently, creativity is a property of an \textit{idea} while diversity is a property of an \textit{idea set.}  We measure two sides of diversity---semantic \textit{divergence} (which we refer to as \textbf{idea diversity}) and semantic \textit{convergence} towards AI ideas (which we refer to as \textbf{AI adoption}). 

To measure idea diversity and AI adoption, we first embed all ideas using SBERT \citep{reimers_sentence-bert_2019}, which are transformer-based embeddings designed for sentences. SBERT excels at capturing semantic similarity \citep{reimers_sentence-bert_2019}. Prior work uses neural embeddings to compute similarity for AUT responses \citep{baten_cues_2021} and other creative tasks \citep{roemmele_inspiration_2021}.

\begin{itemize}
\item[] \textbf{Idea diversity} is the median pairwise cosine distance between idea embeddings in an idea set. As robustness checks, we also measure the mean pairwise distance and average distance to the centroid of a set.  

\item[] \textbf{AI adoption} is the maximum cosine similarity between the embedding of the idea a participant submits and the embeddings of AI examples that the participants see. Following \cite{roemmele_inspiration_2021}, we use the max rather than a measure of central tendency because if a participant is inspired by an idea, it would likely be a \textit{single} idea. As robustness checks, we also measure the mean and median pairwise similarity between the submitted idea and an AI example, but these are noisier measures of adoption. 

\end{itemize}

\section{Related Work} 
\label{Related Research}
Our work bridges three research streams: human-AI co-creation, crowd-sourced creativity, and collective dynamics. AI ideas are scattered amongst human ideas, whether or not we can tell \citep{jakesch_human_2022}. This exposure presumably affects the ideas we create (co-creation). And our ideas presumably affect the ideas \textit{others} create (crowdsourced creativity). While real-world culture is dynamic and evolving, most experiments are not set up to capture evolution (collective dynamics). By employing a large-N sample size and ``multiple-worlds'' setup, we model the complex dynamics of AI influence. In \autoref{detailed_background}, we provide a more detailed overview of how our experimental factors (LLM exposure and disclosure) may affect creativity and diversity of ideas, as well as how attitudinal variables may moderate these effects. Overall, predictions from prior work and theory are conflicting---motivating the present study.

\label{situating}
\subsection{Human-AI Co-Creation}
As AI becomes more creative \citep{miller_artist_2019}, researchers have explored how co-creating with AI affects human creativity. Much of this research explores creative writing with language models \citep{gero_sparks_2022, mirowski_co-writing_2023, lee_coauthor_2022, yang_ai_2022, yuan_wordcraft_2022, roemmele_inspiration_2021, di_fede_idea_2022, hitsuwari_does_2022, gero_ai_2023, mizrahi_coming_2020, gero_metaphoria_2019,padmakumar_does_2024}. While most prior work involves users \textit{actively engaging} with custom systems, our study is concerned with \textit{passive exposure} to off-the-shelf models. The relationship between AI ideas and their effect on human creativity is nuanced. Task-level and attitudinal factors play a role. For example, seeing AI examples influenced outcomes for hard, but not easy, prompts in \citet{roemmele_inspiration_2021}, and \citet{gero_sparks_2022} found the quality of LLM outputs did not correlate with perceived usefulness. Large variance in the perceived usefulness of outputs from co-creation systems~\citep{calderwood_how_2020} suggests human attitudes partially determine the utility of AI creativity aids. We extend this predominantly qualitative work with a large-scale quantitative study.

Several studies, like ours, explored how (post-ChatGPT) generative artificial intelligence affects creativity and diversity. Some studies show that ideating with generative AI decreases diversity \cite{anderson_homogenization_2024,doshi_generative_2024, padmakumar_does_2024}. Other studies suggest generative AI increases individual creativity~\citep{doshi_generative_2024, dellacqua_navigating_2023} while others find mixed effects on creativity~\cite{anderson_homogenization_2024}. Our study offers several additions. First, because of its dynamic design, we test the long-run effects of AI, where ideas feed forward to future participants. Second, our study is concerned with ``passive exposure'': Participants are not told ideas are from AI, and are not instructed to engage with these ideas. By systematically ablating whether AI ideas are disclosed as such, we can explore if the effect of AI ideas depends on knowledge of where the idea is from. Third, we employ a large sample size---which is useful since it provides power for precise estimates of effects and the ability to capture heterogeneity. Fourth, our large sample is comprised of creative professionals and technology-oriented users, two groups relevant to the phenomenon.

\subsubsection{Crowdsourced Creativity}
Crowdsourcing can enhance creative outcomes \citep{yu_cooks_2011, yu_internet-scale_2013, nickerson_crowdsourcing_2010, huang_heteroglossia_2020, siangliulue_toward_2015}. \citet{siangliulue_toward_2015} found that the creativity and diversity of idea sets that participants saw influenced the creativity and diversity of what these participants produced. This supports a main contention of our paper. AI exposure matters because the ideas we see affect the ideas we create. We incorporate elements of crowdsourced creativity, particularly in measuring how creativity and diversity unfold over subsequent generations. 

\subsubsection{Collective Dynamics \& Multiple-Worlds Experiments}
Prior work in complex systems and sociology highlights the importance of studying \textit{collectives} to understand social dynamics \cite{salganik_web-based_2009}. Meanwhile, traditional experiments focus on \textit{individuals}. Identifying the effect of AI ideas on the diversity and evolution human ideas similarly requires an examination of complex systems as opposed to individuals in isolation. Hence, our experimental design draws on the ``multiple-world'' paradigm (e.g., \cite{salganik_experimental_2006, salganik_web-based_2009}): We create multiple, parallel realizations of worlds with and without AI ideas, each evolving independently under controlled conditions. By employing a large-N sample size and many different parallel worlds, we can better understand the collective effects of AI ideas on human ideas. Beyond simple averages, we can also model how AI ideas affect the \textit{evolution} of human ideas.

\subsubsection{Our Contributions}

Our study integrates human-AI co-creation, crowd creativity, and collective dynamics. Traditional co-creation studies typically confound exposure and disclosure effects: when creating with AI, one cannot separate the AI's \textit{content} from the \textit{knowledge} that this content is AI-generated. Our design allows us to estimate these effects independently. Co-creation studies typically employ a small number of specialized participants actively engaged with a system. We are interested in the effects of (1) \textit{passive exposure} on (2) a more general public. For this reason, we adopt a large-scale experimental design---similar to crowd-sourced creativity studies---that lets us estimate effects on the general public rather than specialized users. A key benefit of our large sample size is that we can precisely estimate how participant attitudes affect human-AI outcomes. This is important because attitudes affect human-AI collaboration outcomes~\citep{gero_sparks_2022}. Drawing on the ``multiple-worlds'' paradigm~\cite{salganik_web-based_2009}, our experiment design also lets us understand the effect of AI \textit{over time} since responses feed forward, allowing us to observe differences in \textit{rates of change} between conditions. This method can be used for other experiments to measure the cumulative effect of AI.

\section{Experiment}

\subsection{Summary}
We recruited participants through social media platforms and newsletters. Once participants clicked on the link to the experiment, they were taken to a landing page. In addition to a consent button, that landing page asked several questions related to their self-perceived creativity, attitudes towards AI, and demographics. Following consent, participants completed 5 trials. Each trial required generating a creative use for an object under a specific experimental condition after viewing example ideas---which were our experimental manipulation. Ideas fed forward to future trials such that if a participant was in the \{[\textit{Control}], \textit{tire}\} condition the example ideas the participant saw were the most recent ideas from \textit{prior} participants in the \{[\textit{Control}], \textit{tire}\} condition. See \autoref{fig:clean_screenshot} for experimental conditions. The experiment was approved by our University's IRB. 

\subsection{Stimuli Creation}

\paragraph{Picking AUT Items.}

 Previously, \citet{organisciak_beyond_2023} fine-tuned GPT-3 to predict the creativity of AUT items. Since we use this classifier for our creativity DV, we picked the five items for which the classifier had the highest classifier-human judgment correlations: \textit{tire} (r=0.91), \textit{pants} (r=0.91), \textit{shoe} (r=0.91), \textit{table} (r=0.9), and \textit{bottle} (r=0.88). Our task is highly `in-domain' for the classifier: Participants do the \textit{same task}, using the \textit{same items}, from the classifier training and evaluation. See Appendix \ref{picking_aut_items} for more details on the classifier dataset (``Organisciak Dataset''), training, and performance. 

\paragraph{Model \& Prompting Strategy.}

We generated AI ideas using ChatGPT-3.5 with a zero-shot prompt adapted from \cite{stevenson_putting_2022}. We chose ChatGPT-3.5 for ecological validity as it was the freely available model powering ChatGPT at the time of our study. Zero-shot prompting required no labeled data, making it more practical for typical users. We modified the prompt from \citet{stevenson_putting_2022} to match word counts from the Organisciak Dataset \citep{organisciak_beyond_2023}, preventing easy identification of AI-generated content. Using parameters from \cite{stevenson_putting_2022} and instructions to match human word counts, our modified prompt produced responses similar to human responses in length. These generated ideas served as stimuli for our main experiment. For prompt details and statistical tests, see Appendix \ref{generating_gpt_ideas}.

\begin{figure*}[h!]
    \centering
    \includegraphics[width=0.8\textwidth]{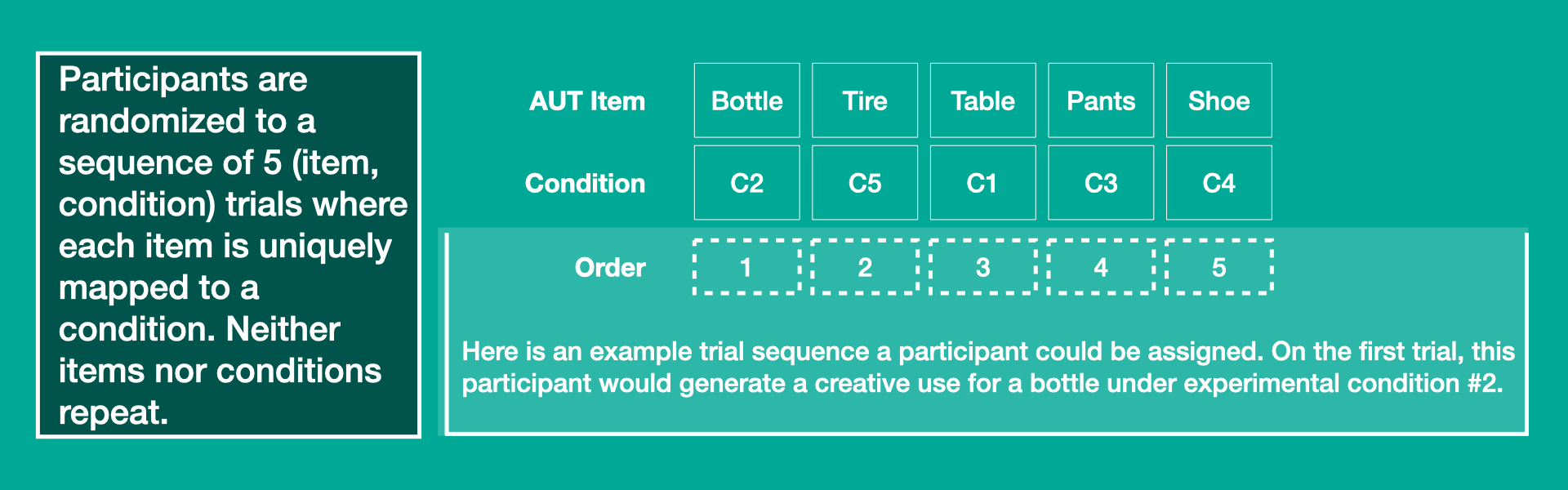}
    \caption{Participants are randomized to a sequence of 5 trials. In each trial, participants generate a creative use for an item under a specific experimental condition. Neither items nor conditions repeat in a 5-trial sequence.}
    \label{fig:trials_small}
\end{figure*}

\subsection{Recruitment Strategy}
We recruited volunteer participants from three sources: (1) Facebook ads, (2) Reddit, and (3) Creative Mornings' weekly newsletter. Creative Mornings\footnote{\url{https://www.creativemornings.com/}}, "the world's largest face-to-face creative community," is an organization for creative professionals that hosts talks and meetups. Creative Mornings agreed to distribute our experiment. We verified all participants were at least 18 years old. While we did not offer monetary compensation, we offered to give participants information about themselves, such as their creativity relative to both humans and AI, and their ability to spot creative ideas. Providing information to participants about themselves is often effective for recruiting volunteers since it makes the task intrinsically rewarding \citep{reinecke_labinthewild_2015}. See Table \ref{sources} for sources and SI 6 for sample characteristics.

We recruited volunteer participants instead of crowdsourced workers for several reasons. First, we wanted participants to be intrinsically motivated since (1) many theories suggest intrinsic motivation helps creativity \citep{mumford_handbook_2017} and (2) we did not want low-quality engagement to confound results (especially since ideas propagate forward). Second, we recruited participants in a targeted manner. In particular, we wanted to target (1) individuals who have a demonstrated interest in technology and (2) those who have a demonstrated interest in creativity. These groups are most relevant to the phenomenon in question. We reached technology-oriented users by posting the experiment in the following subreddits: r/chatgpt, r/InternetIsBeautiful, r/singularity, and r/artificial. We reached creativity-oriented users by posting the experiment in r/writing, r/poetry, and the Creative Mornings newsletter. We also used several `neutral' sources to test the experiment: r/samplesize and Facebook ads. If a participant completed the experiment, then the participant was given a shareable link to their results so they could spread the study.

\subsection{Experiment Procedure}
Once participants clicked on our link, they went to a landing page that included a consent form, task description, and pre-treatment questions. See \autoref{fig:study_description} for the task description. See SI 5 for more details on how participant feedback was calculated.

\begin{figure}[h]
\centering

\fbox{%
\begin{minipage}{0.8\columnwidth}
    \textbf{What you will do:} \\
    We'll show you 5 common items, and you'll come up with creative uses for each item. To spark your imagination, you'll see ideas from prior participants and even from AI (i.e., ChatGPT). You'll be asked to rank these ideas in order of creativity. The ideas you write may be anonymously shown to future participants to spark their imagination. The study takes 3-6 minutes to complete. The goal is to learn about how humans and AI brainstorm.
    \vspace{1.5em}
    
    \textbf{What you will learn:} 
    \begin{itemize}
        \item How creative you are compared to other humans
        \item How creative you are compared to AI
        \item How well you can rank creative ideas 
    \end{itemize}
    \vspace{1.5em}
    We will give you a shareable link with results at the end.
\end{minipage}%
}

\caption{Study description provided to participants.}
\label{fig:study_description}
\end{figure}

\subsubsection{Pre-Treatment Questions.}Participants were asked several pre-treatment questions (\autoref{fig:questions_asked}), including self-perceived creativity, perceived creativity of AI, and attitude towards AI. The third question was from Pew~\cite{nadeem_public_2023, nadeem_how_2022}. We chose the Pew question instead of a longer battery of questions about AI to minimize the response burden. See SI 2 for more details about these questions.

\begin{figure}[h]
\centering
\fbox{%
\begin{minipage}{0.8\columnwidth}
\begin{enumerate}
    \item (required) A slider ranging from 0 to 100 that says `I am more creative than X\% of AI`
    \item (required) A slider ranging from 0 to 100 that says `I am more creative than X\% of Humans`
    \item (required) `Artificial intelligence computer programs are designed to learn tasks that humans typically do. Would you say the increased use of artificial intelligence computer programs in daily life makes you feel...[`More concerned than excited', `More excited than concerned', `Equally excited and concerned']
    \item (optional) What country are you from?
    \item (optional) What is your age?
    \item (optional) What is your gender?
\end{enumerate}
\end{minipage}%
}

\caption{Pre-Treatment Questions}
\label{fig:questions_asked}
\end{figure}

\subsubsection{Randomization}
Participants were assigned a sequence of 5 trials, where each trial was a \{[\textit{condition}], \textit{item}\} pair. For example, one trial might be a creative idea for \textit{pants} in the [\textit{High Exposure}, \textit{Disclosed}] condition. We mapped each AUT item (pants, tire, shoe, bottle, table) to one of the five conditions such that neither conditions nor items repeated in a 5-item sequence. See Figure \ref{fig:trials_small} for a visual explanation.

\subsubsection{Task Instructions}
For each trial (Figure \ref{fig:trial_box}), we asked participants to first rank a list of example ideas by creativity and then submit their own idea. See Appendix \ref{experiment_screenshots} for screenshots. We included the ranking task to ensure engagement with the example ideas. Across conditions, we manipulated two factors: (1) \textit{exposure}---whether AI-generated ideas were included in the example set, and (2) \textit{disclosure}---whether AI ideas were labeled. We used the same prompt for human participants as \citet{stevenson_putting_2022} used with ChatGPT, with a minor modification requesting a single idea (for reasons discussed in \ref{creativity}). This prompt incorporates language aligned with divergent thinking assessment best practices \citep{beaty_forward_2021}. After idea submission, participants received feedback on their idea's uniqueness and their ranking accuracy (screenshots in Appendix \ref{experiment_screenshots}).

\begin{figure}[h]
\centering
\fbox{%
\begin{minipage}{0.8\columnwidth}
    \textbf{Task} \\
    For this task, you will submit a creative use for a [ITEM]. But before submitting your idea, here are some ideas for inspiration. Rank them by creativity. 
    \vspace{1.2em}
    
    \textbf{Rank Previous Ideas} 
    \begin{itemize}
        \item  Rank these ideas in order of creativity, with the most creative use on top. Drag ideas to rank them.
        \item We’ll show you how your rankings compare to rankings from a highly accurate model.
    \end{itemize} 
    
    \vspace{1.2em}
    [SORTABLE EXAMPLE IDEAS HERE] 
    
    \vspace{1.2em}
    \textbf{Submit Your Idea} \\
    Your turn! What is a creative use for a [ITEM]? The goal is to come up with a creative idea, which is an idea that strikes people as clever, unusual, interesting, uncommon, humorous, innovative, or different. List a creative use for a [ITEM].
\end{minipage}%
}
\caption{Task Instructions}
\label{fig:trial_box}
\end{figure}

\subsubsection{Response Chains}

\begin{figure}[h!]
    \centering
    \includegraphics[width=0.5\textwidth]{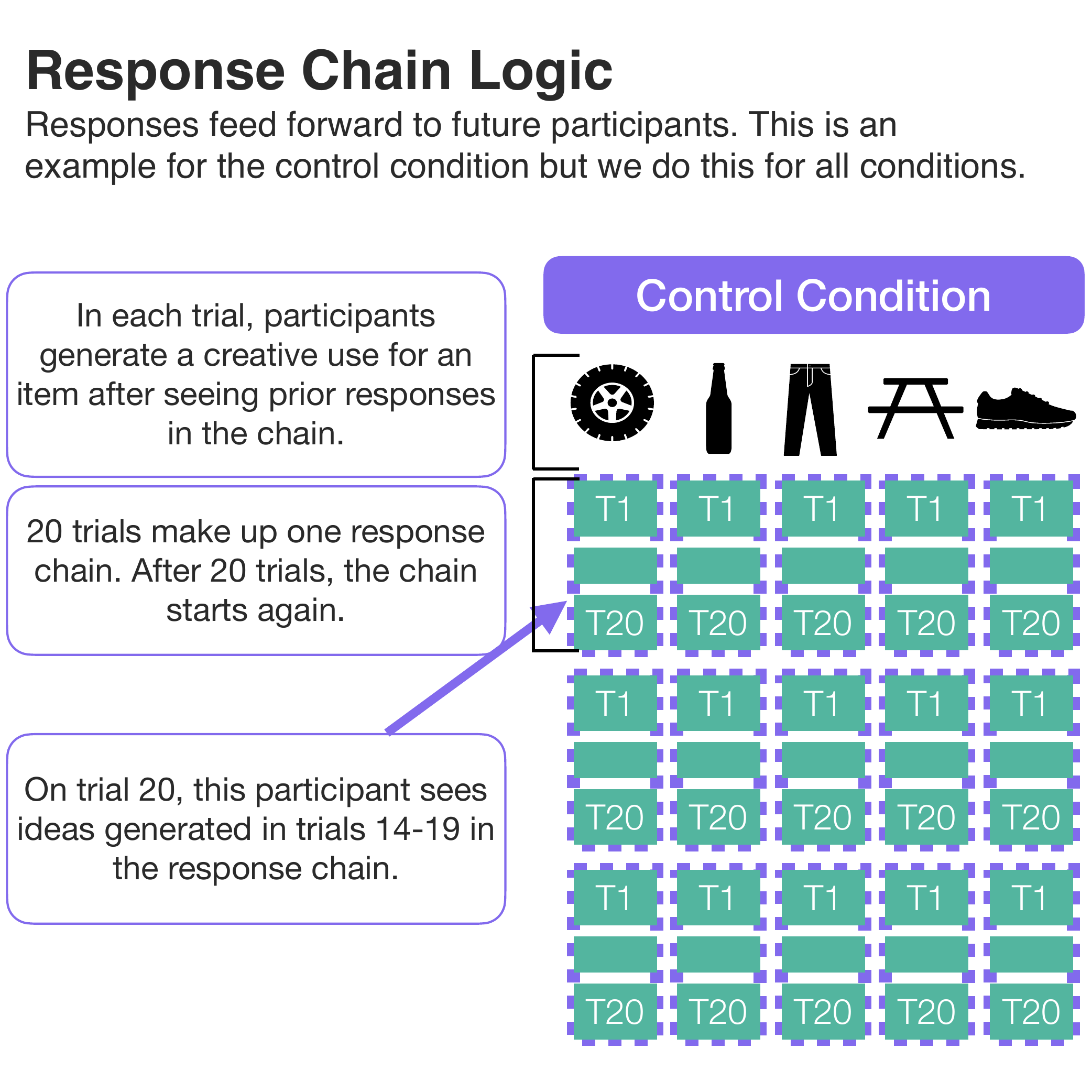}
    \caption{Participants see example ideas from prior participants in the same condition. These `response chains' reset every 20 responses.}
    \label{fig:trials}
\end{figure}

\paragraph{Logic}
The human ideas that participants saw came from prior participants in the same \{[\textit{condition}], \textit{item}\} combination. See Figure \ref{fig:trials}. For instance, if a user was placed in the [\textit{Control}] condition for a \textit{tire}, that user would see six human ideas---the most recent six ideas for a \textit{tire} under the [\textit{Control}] condition. In order to avoid overfitting to a specific idea sequence, we reset this `response chain' every 20 trials. So, the first 20 participants in the \{[\textit{Control}], \textit{tire}\} combination would see each other's ideas, but the chain would reset for the 21st respondent. We use the logic described in this paragraph and Figure \ref{fig:trials} for the human ideas in all conditions. 

Note that because human ideas are propagated at the \{[\textit{condition}], \textit{item}\} level, the human ideas in the [\textit{Control}] condition are `clean' from AI contamination. They were brainstormed after seeing sets of human-only ideas, also from the [\textit{Control}] condition.

We ran seven response chains for each of the 25 (5 items $\times$ 5 conditions) combinations, corresponding to 175 response chains in all and 3500 targeted responses (175 response chains $\times$ 20  trials per chain).

\paragraph{Human Seeds} 
Of course, there is a bootstrapping problem---what human ideas does the \textit{first} person in the \{[\textit{Control}], \textit{tire}\} condition see? The seeds for each \{[\textit{condition}], \textit{item}\} combination came from prior responses from the Organisciak Dataset. That is, Participant 1 for a \{[\textit{Control}], \textit{tire}\} response chain would see 6 seed items. Then Participant 2 in the same response chain would see 5 seed items plus Participant 1's idea (the order of ideas is randomized). Participant 3 would see 4 seed items plus Participant 1 and Participant 2's ideas, etc. We chose a random sample of seeds for each \{[\textit{condition}], \textit{item}\} combination from the Organisciak Dataset. The dataset labeled ideas with gold-standard human ratings of originality. We conducted an ANOVA and found no significant condition-level difference in the originality of the seeds we used.

\section{Outcome Measures}

\begin{table*}[t]
\centering
\caption{Three outcome measures (idea diversity, creativity, AI adoption) across three levels of analysis (local, global, evolution). All ideas embedded using SBERT.}
\label{big_table}
\small
\renewcommand{\arraystretch}{1.2}
\begin{tabular}{p{2.2cm} p{4cm} p{3.2cm} p{4cm}}
\toprule
& \textbf{Local} & \textbf{Global} & \textbf{Evolution} \\
\midrule
\textbf{Creativity} &
How creative is the submitted response? Measured by the prediction from the Organisciak et al. classifier. &
Not applicable &
Does the creativity of submitted ideas change over time? Measured by the slope of trial number (i.e., iteration in response chain) on creativity scores from Organisciak et al. \\
\addlinespace[0.5em]
\textbf{Idea Diversity} &
How different is a participant's response from example responses? Measured by median pairwise semantic distance between ideas a participant sees and their response (SI 8.3). &
How diverse were all participants' ideas in a condition? Measured by median pairwise distance between all submitted ideas in a condition. &
Do ideas become more different from each other as the experiment progresses? We measure the median pairwise distance of submitted ideas at each trial number, then calculate the slope of trial number on idea diversity. \\
\addlinespace[0.5em]
\textbf{AI Adoption} &
How similar is a participant's response to AI example responses? Measured by the maximum semantic distance between a participant's response and AI examples. &
Not applicable &
Not applicable \\
\bottomrule
\end{tabular}
\end{table*}

We have three outcome measures (idea diversity, creativity, and AI adoption) and three levels of analysis (local, evolution, and global). See Table \ref{big_table}. The \textit{local} level measures outcomes at the level of an individual trial (e.g., how a submitted response relates to example responses). The \textit{evolution} level measures the rate of change of outcome variables with respect to the trial number in the response chain (i.e., experiment iteration). The \textit{global} level compares all submitted responses in a condition to each other. For all pairwise comparisons, we use a Holm-Bonferroni adjustment for multiple comparisons. For idea diversity and AI adoption, we scale the dependent variable (cosine distance or cosine similarity, respectively) by 100 for easier interpretation. See Appendix \ref{appendix_outcomes} for more methodological details.

\section{Results}

\subsection{Recruited Participants}

We received over 3,000 responses from 48 countries. See SI 6 for sample characteristics. Out of a total of five trials, participants finished four trials on average (Table \ref{overall_stats}), suggesting the experiment was engaging. Most participants came from the Creative Mornings newsletter or r/InternetIsBeautiful (Table \ref{sources} for source counts and categorization). Of the non-missing responses, the sample was 50\% women, 43\% men, 4\% non-binary, 3\% not disclosed, 1\% self-described.  The mean age was 34.92 (SD = 10.86). Regarding AI, the sample was 48\% neutral, 28\% excited, 24\% concerned. Participants said they were more creative than 57.86\% (SD = 26.66) of AI and 58.67\% (SD = 23.65) of humans. See SI 6 for kernel density plots. 

\begin{table}[h!]
\centering
\caption{Summary Statistics of Experiment}
\label{overall_stats}
\begin{tabular}{lr}
\toprule
{} &    Value \\
\midrule
Unique Countries          &    48 \\
Total Responses           &  3414 \\
Unique Participants       &   844 \\
Avg Responses/Participant &     4.05 \\
Avg Duration/Response (Seconds)     &   144.31 \\
\bottomrule
\end{tabular}
\end{table}

\begin{table*}[h!]
\centering
\caption{Sources of participants and trials. For analysis, we categorized each source into a higher-level interest group (technology, creativity, neutral).}
\label{sources}
\begin{tabular}{llll}
\toprule
Interest Group &                       source & Participants (N, \% of total) & Trials (N, \% of total) \\
\midrule
      creative & Creative Mornings newsletter &                  343 (40.6\%) &           1470 (43.1\%) \\
    technology &        r/InternetIsBeautiful &                  298 (35.3\%) &           1115 (32.7\%) \\
       neutral &                 r/samplesize &                   94 (11.1\%) &            389 (11.4\%) \\
       neutral &                        share &                    61 (7.2\%) &             250 (7.3\%) \\
    technology &                    r/chatgpt &                    19 (2.3\%) &              79 (2.3\%) \\
      creative &                    r/writing &                     7 (0.8\%) &              30 (0.9\%) \\
       neutral &                        other &                     6 (0.7\%) &              22 (0.6\%) \\
    technology &                r/singularity &                     6 (0.7\%) &              13 (0.4\%) \\
    technology &                 r/artificial &                     5 (0.6\%) &              24 (0.7\%) \\
      creative &                     r/poetry &                     3 (0.4\%) &              15 (0.4\%) \\
       neutral &                     facebook &                     2 (0.2\%) &               7 (0.2\%) \\
\bottomrule
\end{tabular}
\end{table*}

\begin{figure}[h!]
    \centering
    \includegraphics[width=1\columnwidth]{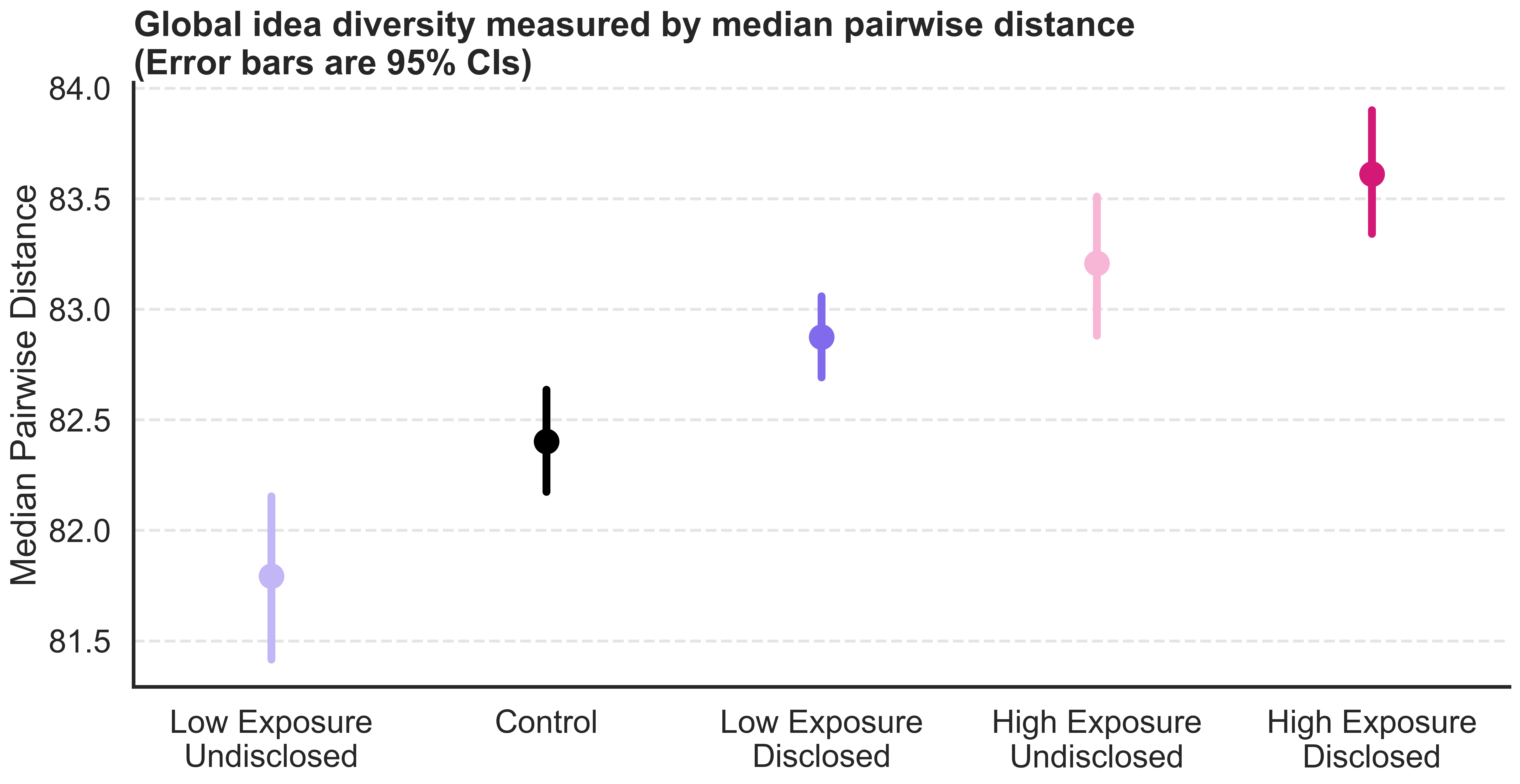}
    \caption{Median pairwise distance of submitted ideas in a condition from Monte Carlo analysis. There was more global diversity of ideas in both high AI exposure conditions than in the control condition.}
    \label{fig:global_diversity_plot}
\end{figure}

\begin{figure}[h!]
    \centering
    \includegraphics[width=1\columnwidth]{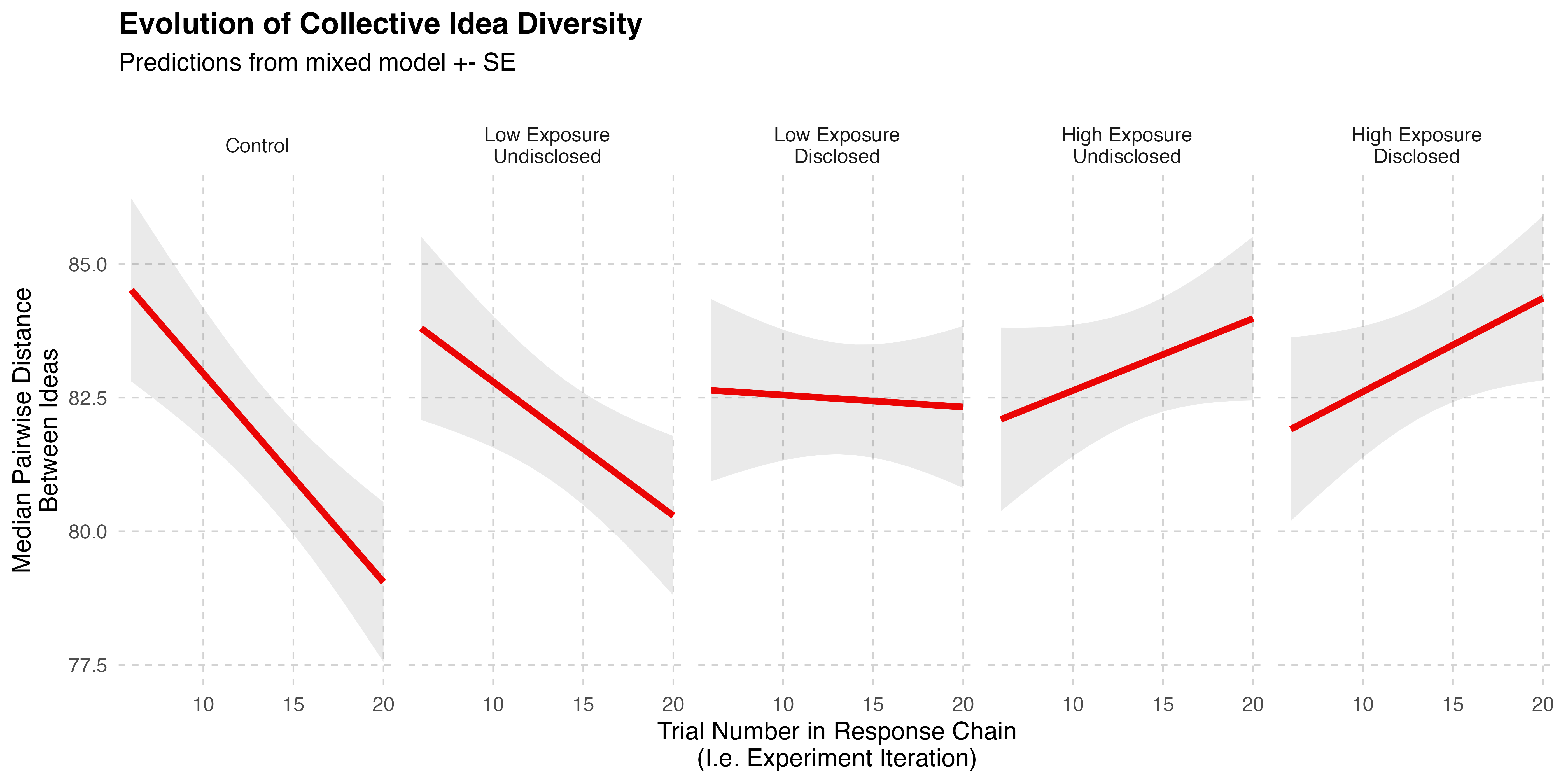}
    \caption{High exposure to AI ideas increased the rate of change in idea diversity. `Idea diversity' was measured as the cosine distance between submitted ideas.}
    \label{fig:evo_diversity}
\end{figure}

\subsection{High AI exposure increases collective diversity} Intuitively, global idea diversity measures how different submitted ideas are in each condition. We measured global idea diversity by looking at the median pairwise distance between neural embeddings of the pool of ideas that were submitted by participants in each condition (Appendix \ref{global_diversity}). As a non-parametric measure of effect size, we calculate Cliff’s Delta ($\delta$) of these pairwise differences, which ranges from -1 to 1. A value of 0 indicates no difference between the two conditions, +1 indicates values from the first condition are always larger, and -1 indicates the opposite.

We find (\autoref{fig:global_diversity_plot}) that both the [\textit{High Exposure, Disclosed}] ($ \delta = 0.31; \text{p}=0.001)$ and [\textit{High Exposure, Undisclosed}] $(\delta = 0.26; \text{p}=0.001)$ conditions had more global idea diversity than the control condition. But of the low exposure conditions, only the [\textit{Low Exposure, Undisclosed}] $(\delta = 0.11; \text{p}=0.001)$ condition had higher global diversity than the control condition, with a much smaller effect size than the high exposure conditions. P-values were computed with permutation tests incorporating a conservative adjustment \cite{ojala_permutation_2009} (adding 1 to numerator and denominator) and a Holm-Bonferroni correction for multiple comparisons. As robustness checks, we ran this analysis with alternative measures of idea diversity (SI 8) and the results are directionally similar. We report results for the median pairwise distance as it provides the most \textit{conservative} estimate.

\subsection{High AI exposure increases the speed at which idea diversity develops} We measured the evolution of diversity by looking at the diversity (Appendix \ref{evo_outcomes}) of submitted ideas at each iteration in the experiment and seeing if---in some conditions---diversity developed at a faster rate. We find that high AI exposure increases not only the average levels (\autoref{fig:global_diversity_plot}) of collective idea diversity but also the rate of change in idea diversity (\autoref{fig:evo_diversity}). As with global idea diversity, different metrics yielded similar regression coefficients (SI 8.3). In the control condition, idea diversity decreased over trials ($\beta = -0.39 $, $t(349) =  -2.23 $, 95\% CI = $[-0.73,  -0.05]$, $p=0.03$). That is, submitted ideas were becoming more similar to each other as the experiment went on. Relative to the control condition, however, the slope of idea diversity with respect to trial number was more positive for the [\textit{High Exposure, Undisclosed}] condition ($\beta =  0.53 $, $t(349) =  2.2 $,  95\% CI = $[ 0.06, 0.99 ]$, $p=0.03$) and the [\textit{High Exposure, Disclosed}] condition ($\beta =  0.57 $, $t(349) =  2.37 $,  95\% CI = $[0.1, 1.03 ]$, $p=0.02$). The rate of change in idea diversity for the low AI exposure conditions did not differ from the rate of change in the control condition. Thus, we conclude that high exposure to AI ideas increased the rate of idea diversity relative to the no-AI, control condition.

\subsection{Seeing AI ideas does not affect creativity} 
Conditions did not differ by creativity of ideas, $F(4,19.86)=0.12,p=0.97$. We used the \textit{emmeans}\footnote{https://cran.r-project.org/web/packages/emmeans/index.html} joint-tests function to assess main effects of condition after a mixed model (Appendix \ref{local_level} for modeling details; SI 9 for regression results). Similarly, no individual treatment condition coefficient differed from zero where the baseline was the control condition. We also tested whether the \textit{evolution} of creativity differed by condition. We conducted a likelihood ratio test on whether adding a [trial number (as an integer) X condition] interaction would improve the fit of a model. The likelihood ratio test indicated that allowing for this temporal interaction \textit{did not} significantly improve the model fit $(\chi^2 (4) = 6.52, p=0.16)$---suggesting conditions did not differ on average \textit{or} with respect to the temporal evolution of creativity. All results are based on mixed models (Appendix \ref{local_level}) with crossed and nested intercepts accounting for non-independence in the experiment.

\subsection{How AI Adoption differs depending on task and self-perceived creativity} 
As a secondary outcome to creativity and diversity, we measured whether AI adoption (the extent to which people submit similar ideas to AI examples) differs depending on the task and person. Prior work found that people may rely more on AI for more difficult tasks \citep{roemmele_inspiration_2021, goddard_automation_2014}. Other work posited that attitudes towards a technology are a major factor when studying human-AI collaboration. \citep{gero_sparks_2022}. We measured three attitudinal variables: Attitude toward AI, self-perceived creativity, and perceived creativity of AI. For our individual-level model of adoption, we only included variables as moderators if they significantly improved the fit of a no-moderator, baseline model (Appendix \ref{local_level}). We operationalized AI adoption as the maximum similarity (following \citet{roemmele_inspiration_2021}) between a submitted idea and AI example ideas a participant saw. The logic for using the maximum is that insofar as a person is influenced by an idea, it would likely be a single one \citep{roemmele_inspiration_2021}.

\begin{figure}[h!]
    \centering
    \includegraphics[width=1\columnwidth]{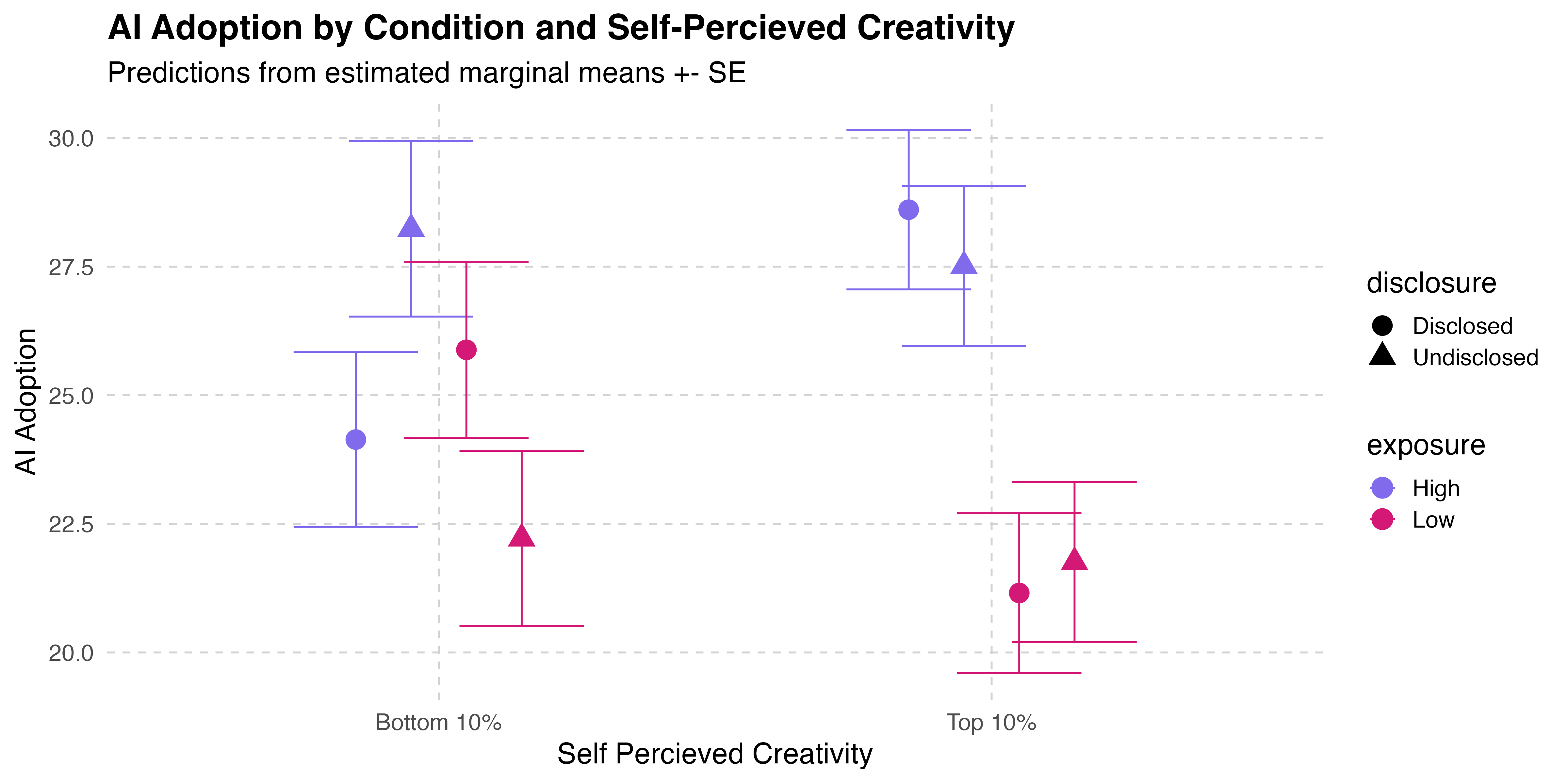}
    \caption{Higher creativity participants adopted AI ideas based on exposure and were not influenced by AI disclosure (i.e., whether AI ideas were labeled as such). `AI Adoption' was measured by the similarity of a submitted idea to AI example ideas.}
    \label{fig:create_adopt}
\end{figure}

\begin{figure}[h!]
    \centering
    \includegraphics[width=1\columnwidth]{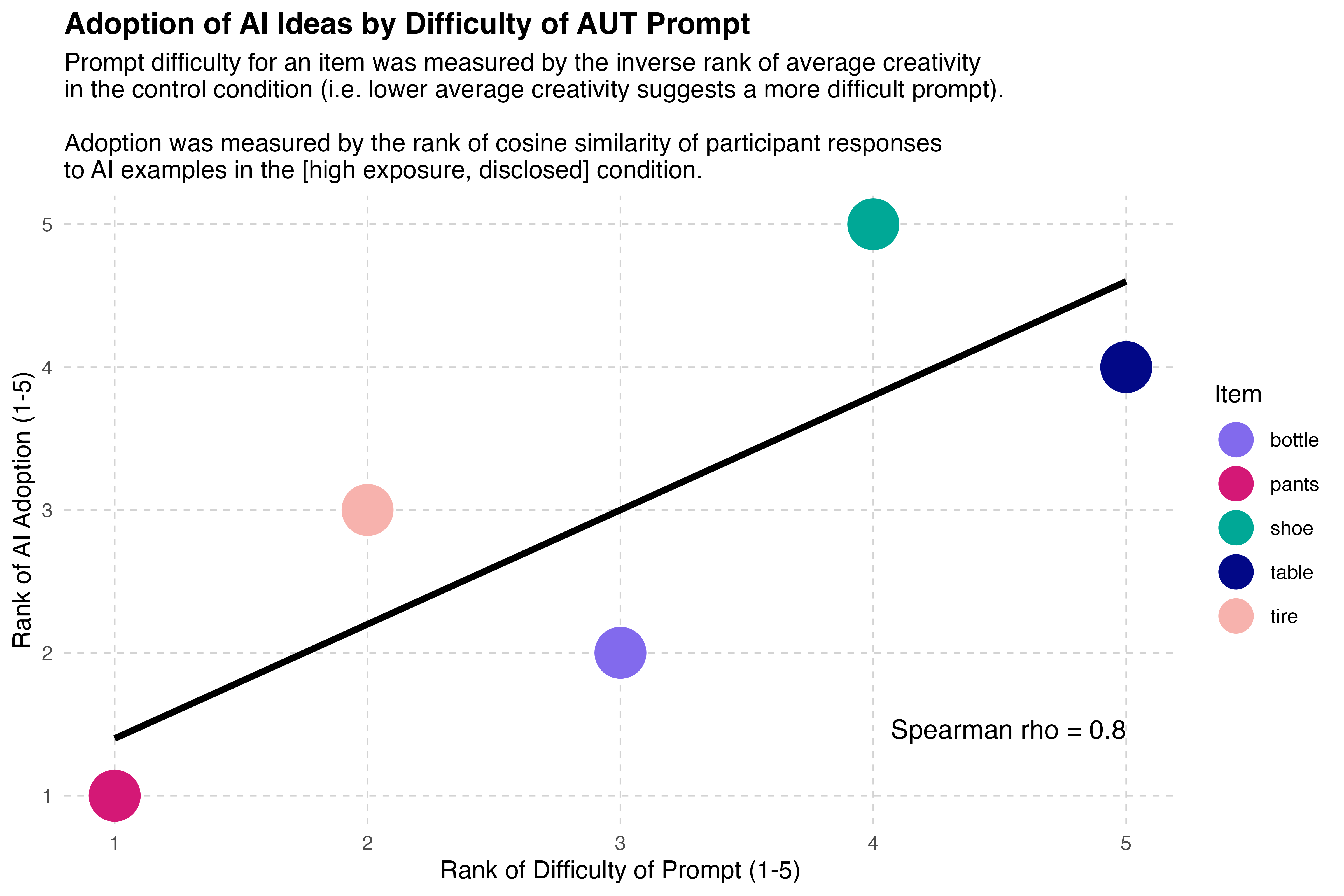}
    \caption{When AI ideas were disclosed as such, participants were more likely to adopt AI ideas for more difficult prompts.}
    \label{fig:adopt_difficult}
\end{figure}

\subsubsection{People who identify as creative are less influenced by AI disclosure} 
There was a significant interaction between self-perceived human creativity and the [\textit{High Exposure, Disclosed}] condition ($\beta =  0.11 $, $t(2588) =  3.93 $,  95\% CI = $[0.06, 0.17 ], p = 0.0001$; SI 10). To probe this interaction, we used our model to predict the estimated marginal mean AI adoption by condition, for both the top 10\% and bottom 10\% of participants by self-perceived creativity (\autoref{fig:create_adopt}).  For high-creativity participants, the effect of exposure on adoption does not differ by whether AI ideas are disclosed as such: There is no difference in adoption between ([\textit{High Exposure, Undisclosed}] - [\textit{Low Exposure, Undisclosed}]) and ([\textit{High Exposure, Disclosed}] - [\textit{Low Exposure, Disclosed}]), $\Delta = -1.69, d=-0.14, p = 0.59$. But for low-creativity participants, the difference in adoption for the \textit{undisclosed} conditions ([\textit{High Exposure, Undisclosed}] - [\textit{Low Exposure, Undisclosed}]) was larger than the equivalent difference in adoption for \textit{disclosed} conditions ([\textit{High Exposure, Disclosed}] - [\textit{Low Exposure, Disclosed}]), $\Delta = 7.77, d=0.65, p=0.03$. In short, disclosing ideas as from AI reduced the effect of AI exposure on adoption for lower creativity participants but not higher creativity participants. P-values on post-hoc estimated marginal means were Holm-Bonferroni adjusted for multiple comparisons.

\subsubsection{Participants may adopt AI ideas more for harder prompts} We find that when AI ideas are disclosed, participants are more likely to adopt the ideas of AI for difficult AUT prompts (\autoref{fig:adopt_difficult}). The rank-order correlation between difficulty and adoption was $\rho=0.8$ for the [\textit{High Exposure, Disclosed}] condition but only $\rho = 0.3$ for the [\textit{High Exposure, Undisclosed}] condition. That is, when people were told ideas were from AI, they were more likely to adopt AI ideas if the prompt was difficult. But since we employed only five items, this finding is speculative. We encourage future work to follow up on this finding.

\section{Discussion}
Against the backdrop of widespread LLM exposure, we asked: How does exposure to ideas generated by LLMs affect the creativity, diversity, and evolution of human ideas? To answer this, we conducted a large-scale experiment where participants submitted ideas in response to the Alternate Uses Task (a measure of creativity where people brainstorm novel uses of an item) after seeing a set of example ideas. The human examples were from prior participants in the same experimental condition, and the AI ideas were from ChatGPT. The evolving aspect of our experiment, that ideas in a condition feed forward to subsequent trials in that condition, captures the fact that ideas build upon prior ideas. This lets us approximate the cultural evolution effects of AI. Here are three takeaways from our experiment. 

\subsection{AI makes ideas different but not better}
Most notably, exposure to AI ideas did not, on average, make human ideas any `better' or `worse' (by creativity). Our high-powered, null finding around average creativity by condition can inform debates about the effect of AI ideas on individual human creativity. Maybe there is little effect. Of course, our experiment is measuring just a single task. But these results suggest that perhaps both worry and optimism around the effect of AI ideas on individual human creativity should be tempered.

Our null finding around creativity contrasts with some prior work suggesting human-AI co-creation enhances the quality of creative outputs \citep{mizrahi_coming_2020, yuan_wordcraft_2022, roemmele_inspiration_2021, hitsuwari_does_2022}. But our study differs from prior studies in its aim and design: We test \textit{passive} exposure to \textit{off-the-shelf} LLMs---not \textit{active engagement} with \textit{optimized-for-creativity} AI aides. The latter is useful for understanding how AI \textit{could} affect creativity. But we aim to approximate how ordinary, existing, and pervasive AI tools \textit{do} affect the creativity of ideas. At least for this task, we find no evidence of such an effect.

On the other hand, the presence of AI ideas increased the \textit{diversity} of human ideas. This is consistent with work suggesting collaborating with AI leads to more diverse or unexpected outputs \citep{yang_ai_2022, osone_buncho_2021, lee_coauthor_2022, branch_collaborative_2021, gero_metaphoria_2019} and inconsistent with other work that finds collaborating with LLMs decreases diversity \cite{anderson_homogenization_2024, doshi_generative_2024, dellacqua_navigating_2023}. But we highlight that our study is testing \textit{passive exposure} to AI ideas and not \textit{active engagement} with AI ideas. Our setup---passive exposure to AI ideas, scattered amongst human ones---maps onto how many users now experience AI ideas. Hence, it may be that active engagement with LLMs decreases content diversity but simply seeing these ideas as ``sparks'' \cite{gero_sparks_2022} \textit{increases} content diversity. And because many more people may be passively exposed to LLM outputs than actively engaging with LLMs, the effect of passive exposure is important to understand. 

Another key difference between our study and many similar studies is the brainstorming context. Prior studies typically examine individual participants working directly with AI systems. Our study simulates group brainstorming where participants see ideas from prior group members. In group contexts, AI ideas may serve as additional ``voices'' that prevent convergence in human-only groups. This suggests AI's effect on ideas may depend on whether it's used for individual ideation versus collective brainstorming. But since ideas we see affect the ideas we create~\cite{nijstad_how_2006}, there is an element of collective (in a broad sense) brainstorming even to individual brainstorming.  

Crucially, high AI exposure increased both average amounts of diversity \textit{and} rates of change in idea diversity. The latter result is especially important. Small differences in rates of change can accumulate to create large aggregate differences over time. This points to a dynamic that may be missed by static experimental designs. Future work---both simulations and dynamic experiments---can explore the implications of this increase in collective idea diversity unaccompanied by an increase in average individual creativity. For instance, can this dynamic generate ``innovation''? 

Our finding around the evolution of diversity (Figure \ref{fig:evo_diversity}) is instructive and points to a likely mechanism: Seeing other people's ideas reduced idea diversity in the control condition over time, as successive participants were ``converging'' on a particular idea sequence. But then injecting AI ideas into the example set increased the diversity of submitted responses by partially ``resetting'' this convergence. This suggests off-the-shelf LLMs may be viewed as extensive, accessible idea banks that can counter groupthink in collaborative settings. Future research should examine the dynamics and boundary conditions of this potential mechanism's effectiveness. 

\subsection{High creativity people are less influenced by the source label of ideas} 
For participants who viewed themselves as highly creative, adoption was driven by exposure and not disclosure (i.e., knowing ideas were from AI). But for lower-creativity participants, knowing the source of an idea \textit{did} affect the adoption of that idea. Perhaps people high in self-reported creativity relied less on source cues when adopting ideas because they were more confident in their ability to judge an idea's creative merit. Regardless, our results suggest that (self-reported) creative people will adopt ideas on the basis of their content. Knowing the source does not matter. In a world where humans have difficulty distinguishing if the content was AI-generated \citep{jakesch_human_2022}, these findings suggest people high in (self-reported) creativity may not be `duped' into adopting AI ideas. 

\subsection{Participants may adopt AI ideas when the task is difficult}
When AI ideas were labeled, participants were more likely to adopt AI ideas for difficult prompts rather than easy prompts. This finding is similar to what \citet{roemmele_inspiration_2021} observed, where seeing AI examples only influenced creative output when the task was difficult. Both findings are consistent with a theoretical account of task difficulty being associated with increased reliance on automation \citep{goddard_automation_2014}. However, we emphasize that this finding should be taken as speculative (due to the small number of items) and not conclusive. We encourage future work to test whether users adopt AI ideas for more challenging creative tasks. 

If it is the case that users turn to AI for difficult rather than trivial tasks, this would have several implications. While AI can potentially augment humans when they are stuck, researchers raised concerns over `model collapse' \citep{shumailov_curse_2023}---the deteriorating performance of LLMs when trained on their outputs. If reliance on AI for creative tasks becomes routine, this may contribute to model collapse. 

\subsection{Limitations And Future Work}
Our study has limitations that should be noted and can inform future work. First, we measured the effect of AI ideas for a \textit{single task}. We chose this task because it is one of the most common creativity tasks \citep{abraham_gender_2016}. But future work could explore other kinds of tasks---especially more complex ones. 

Second, we had to operationalize `AI ideas' in some concrete way (among a large array of choices), and future work can use a similar design but operationalize `AI ideas' in a different way. We chose a commonly-used LLM at the time of the study, and our prompt was driven by ecological validity and prior work: We used a zero-shot prompt because that is what \textit{users} would likely use, and the specific prompt we used was derived from prior research. We chose not to vary prompts because we didn't want to increase the complexity of an already complex experiment. Future work could explore whether different AI elicitation procedures yield different results. This could involve different models, prompts, LLM-based ideation systems (e.g., ~\cite{rick_supermind_2023, pu_ideasynth_2025}), or ``simulated social ensembles''~\cite{ashkinaze_plurals_2025}. A specific design choice we made is that LLMs were not shown any example stimuli, but humans were. We did this to (A) reflect real-life LLM usage (which is often zero-shot, not few-shot) and (B) conceptually represent LLMs as a `bank' of ideas that can break up convergence during group brainstorming. However, this specific operationalization leaves open the question of how LLMs versus humans may come up with different ideas if they were given the same example stimuli. 

Third, our finding about AI adoption and task difficulty is based on only five AUT items (tire, pants, shoe, table, bottle) and should be considered speculative, not definitive. We encourage future work to explore this hypothesis further. Fourth, future work should test if alternative classifiers or ways of conceiving variables yield different results. For idea diversity and AI adoption, we addressed this problem by showing that conceptually similar ways of measuring variables yielded qualitatively similar results. For our creativity measure, we used a highly accurate classifier (correlation with human judgments greater than $r=0.88$ for AUT items we used) trained for this \textit{exact task}, for these \textit{exact AUT items} (Appendix \ref{picking_aut_items}). But of course, all models have some error, and future work based on this model propagates these errors. Incidentally, human judges of creativity only correlate with other human judges at $r=0.88$ \citep{organisciak_beyond_2023}, suggesting the classifier we used may be approaching `the approximate ceiling at which we could expect a model to correlate with human judgments' \citep[pg. 11]{organisciak_beyond_2023} of creativity. Fifth, the classifier measures one facet of creativity: originality. Future work can also explore whether AI ideas have different effects on other facets of creativity such as fluency (how many ideas) or elaboration (how detailed an idea is). Sixth, we employed a non-representative sample of technology-interested users and creative professionals. While these two groups were most relevant to the phenomenon in question, our sample also limits generalizability. Seventh, our design captures some aspects of human-AI cultural evolution but not others; it is a stylized model that does not capture all the complexity of how AI can shape collective behavior~\cite{farrell_large_2025, brinkmann_machine_2023}. For example, future work could explore more complex transmission scenarios.

Despite these limitations, our work offers a large-scale, dynamic account of how seeing LLM ideas affects collective thought. This work opens the door to more dynamic experiments exploring the compounding effects of ``AI in the culture loop''.

\subsection{Conclusion: Passive exposure to AI ideas affects collective thought}
Our experiment suggests that passive exposure to AI ideas---the kind of passive exposure we are now \textit{inundated with} in a post-ChatGPT era---does affect collective thought. Even small effects are meaningful. This exposure is both pervasive and growing. But the effects of AI ideas are nuanced. Seeing AI ideas did not increase individual creativity, though it did increase collective diversity. The effects of AI ideas vary across individuals and tasks. There is still much to learn. We hope our study inspires more research on how passive exposure to AI ideas affects collective thought via the use of multiple-worlds experiments.

\begin{acks}
We thank several of our colleagues for feedback, including: Joyce Chai, Harman Kaur, and Misha Teplitskiy. Apart from where described in the manuscript, we used generative AI for coding completions and light writing edits. 
\end{acks}

\bibliographystyle{ACM-Reference-Format}
\bibliography{smaller_bib}

\appendix

\section{Stimuli Construction} \label{stimuli}
\subsection{Choosing AUT Items} \label{picking_aut_items}
We had to choose a selection of items that people would brainstorm creative uses for. We chose five items for which the creativity classifier that we used had the highest accuracy. Previously, \citet{organisciak_beyond_2022} fine-tuned GPT-3 Davinci to predict the creativity of AUT items. We use a dataset from \citet{organisciak_beyond_2022} containing 20,121 responses from 2,025 participants, across 21 distinct AUT items and nine distinct studies \citep{organisciak_beyond_2022}.\footnote{We obtained an early version of this dataset by direct correspondence with Dr. Organisciak on February 23, 2023; the code that Dr. Organisciak used to generate this dataset is available at \url{https://github.com/massivetexts/llm_aut_study/blob/main/notebooks/Process_AUT_GT.ipynb}} Each response was graded for creativity by humans and normalized to a scale of 1-5. Then \cite{organisciak_beyond_2022} fine-tuned GPT-3 Davinci to predict creativity scores. Here, fine-tuning involves providing \{\textit{Input} (an AUT response), \textit{Output} (human rating)\} pairs to a pre-trained LLM. Then the LLM adjusts its parameters to produce a similar output given an input, proxying human judgements. Overall, the fine-tuned GPT-3 classifier had a correlation\footnote{We obtained scores for this classifier by downloading the zip file from (\url{https://github.com/massivetexts/llm_aut_study/blob/main/results/evaluation.zip}), then navigating to \texttt{gt\_main2/gpt-ft-davinci-1.csv}} of $r=0.81$ with human judgment. Accuracy varied by item (SI 1). For our experiment, we picked the five items for which the classifier had the highest accuracy: \textit{tire} (r=0.91), \textit{pants} (r=0.91), \textit{shoe} (r=0.91), \textit{table} (r=0.9), and \textit{bottle} (r=0.88).  Our task is ``in-domain'' for the classifier since we ask participants to do the same task for the same items the classifier was trained on.

\subsection{Generating GPT Ideas} 
\label{generating_gpt_ideas}

\paragraph{Prompt Construction} Our specific zero-shot prompt was informed by prior work on LLMs and creativity. \cite{stevenson_putting_2022} administered the AUT to GPT-3 through a zero-shot prompt. However, this prompt generated much wordier responses $(M = 24.1, SD = 8.8)$ than the human responses in the Organisciak Dataset $(M= 4.6, SD = 5)$. Such a discrepancy would alert participants to what was AI vs. human generated, which would nullify the disclosure factor (whether the source of an idea is disclosed). Hence, we appended a request (Figure \ref{fig:zero_shot_prompt}) to use roughly the same number of words (5) as the average human response.

\paragraph{Word Length Prompt Experiment} We conducted a Monte Carlo experiment to confirm our modified zero-shot prompt resulted in responses with a similar word length to humans. Using parameters from \citet{stevenson_putting_2022}'s experiments: For $n=1000$ trials, we fixed presence penalty and frequency penalty at 1, randomly chose a temperature (higher values lead to more randomness) in [0.65, 0.7, 0.75, 0.8], and randomly chose one of our 5 AUT items. For each trial, ChatGPT generated five ideas. The modified prompt resulted in responses with an average word length (M=4.44, SD = 1.34) much closer to human responses (M=4.56, SD = 4.97) than the original zero-shot prompt (M = 25.38, SD = 8.55). A permutation test (more details below) further shows that this difference was significant at $p<0.001$. We used the ideas generated by our modified zero-shot prompt as stimuli for the main experiment. 

We conducted a permutation test of whether the limited-length zero-shot prompt (`ZeroShotLimit') yielded more similar word counts to the human responses than the original zero-shot prompt (`ZeroShot') from \citep{stevenson_putting_2022}. Note that permutation tests are non-parametric, so they are robust to (e.g.) violations of normality. The observed difference is computed as: 

$$T^{obs} =$$ 
$$ |Avg(\text{ZeroShot}) - Avg(\text{HumanResponses})| -$$
$$ |Avg(\text{ZeroShotLimit}) - Avg(\text{HumanResponses})|$$ where $Avg$ is the average word count of a response and the human responses are from \cite{organisciak_beyond_2022}.

Under the null hypothesis, we assume prompts do not differ in respective word counts, so we can swap the labels of `ZeroShot' and `ZeroShotLimit'. To simulate draws from this null distribution, we permuted the labels of the source column 10000 times, each time calculating a new value for the difference of differences defined in $T$. The p-value is the proportion of permutations where the null $T$ is greater than or equal to the observed difference, $T^{obs}$. As in \cite{ojala_permutation_2009}, we apply a conservative adjustment and add $1$ to the numerator and denominator (this means we never get a p-value of 0). 

$$\text{p} = \frac{[\sum_{i=1}^{n} I (T^{null} >= T^{obs})]+1}{n+1}$$

This p-value indicates the likelihood of observing a difference as extreme as our observed difference if the null hypothesis---that prompts did not differ in word counts---is true. We conclude the observed difference is significant. 

\begin{figure}
    \centering
    \MakePrompt[0.5]{What are some creative uses for a [OBJECT]? The goal is to come up with creative ideas, which are ideas that strike people as clever, unusual, interesting, uncommon, humorous, innovative, or different. List creative uses for a [OBJECT]. \textbf{Make sure each response is [MEAN HUMAN WORDS] words.}}{}
    \caption{To generate AUT ideas, we used the zero-shot prompt from \cite{stevenson_putting_2022} with an additional instruction at the end to match the mean length of human responses from prior work.}
    \label{fig:zero_shot_prompt}
\end{figure}

\section{Outcome Measures}
\label{appendix_outcomes}

\subsection{Local Level}
\label{local_level}

Outcomes at the local level---the level of an individual trial---are useful for two reasons. First, this level shows how a participant's response relates to the examples they see. Second, this level lets us model whether individual differences moderate the effect of either disclosure or exposure. For each of our local outcomes, we have a baseline model that uses crossed random intercepts to account for the multilevel structure of the experiment. The first random intercept is for participants, accounting for clustering due to repeated measures. This random intercept is then crossed with a second random intercept for response chains, which we nest inside of  \{[\textit{condition}], \textit{item}\} combinations. In R syntax, the random effect structure was \texttt{... + (1|ParticipantID) +} \\
\texttt{(1|ItemCondition/ResponseChainID).} Models were fit in the \textit{lme4} R package. We computed profile likelihood confidence intervals for coefficients using the \textit{confint} R package. We used estimated marginal means (\textit{emmeans} R package) to conduct model-adjusted F-tests, linear contrasts, predictions, and pairwise comparisons. We apply Holm-Bonferroni adjustments to pairwise comparison p-values. Our baseline `local' model is:

\begin{align}
\text{variable}_{ijk} &= \beta_0 + \beta_1 \text{condition}_{j} + \beta_2 \text{CreativityHuman}_{i} + \nonumber \\
&\quad \beta_3 \text{AiRelCreate}_{i} + \beta_4 \text{AiFeeling}_{i} + \nonumber \\
&\quad \beta_5 \text{InterestGroup}_{i} + \beta_6 \text{ConditionOrder}_{ijk} + \nonumber \\
&\quad \beta_7 \text{LogDuration}_{ijk} + \beta_8 \text{nSeedsPresent}_{ijk} + \nonumber \\
&\quad \beta_9 \text{TrialNo}_{jk} + u_{0i} + v_{0jk} + e_{ijk}
\end{align}

where

 \begin{itemize}
    \item \(i\) indexes participants, \(j\) indexes item-condition combinations, \(k\) indexes response chains.   
        \item \textbf{\text{CreativityHuman}} is self-perceived creativity relative to AI. 
        \item \textbf{\text{AiRelCreate}} is constructed as (self-perceived creativity to humans) - (self-perceived creativity to AI). Note that this is an implicit measure of AI's (perceived) creativity relative to humans. For example, if you say you are more creative than 40\% of humans and 60\% of AI, then AiRelCreate = -20, as the implicit belief is that AI is \textit{less creative} (-20 percentile points) than humans. Conversely, if you say you are more creative than 50\% of humans but 30\% of AI then the implicit belief is that AI is \textit{more creative} (50\% - 30\% = +20) than humans. 
        \item \textbf{\text{AiFeeling}} refers to the AI sentiment question.
        \item \textbf{\text{InterestGroup}} maps each source of the experiment to categories: creative, neutral, or technology. These categories are described in Table \ref{sources}.
        \item \textbf{\text{ConditionOrder}} denotes the sequence in which the participant was assigned to complete the trial (e.g., the 1st time a participant took part, etc.).
        \item \textbf{\text{LogDuration}} is the natural logarithm of the time (in seconds) a participant spent before submitting their answer.
       \item \textbf{\text{nSeedsPresent}} controls for the number of examples the participant saw that were seed ideas from the Organisciak Dataset. 
        \item \textbf{\text{TrialNo}} indicates the trial number within a specific response chain. For example: the 18th response for \{[\textit{Control}], \textit{tire}, response chain 5\}
\end{itemize}

We balanced interest in testing experimental hypotheses that conditions differed by subgroups with caution around model overfitting. We considered interactions between the treatment condition and four potential moderators: self-perceived human creativity ($CreativityHuman$), AI - Human creativity ($AiRelCreate$), feeling towards AI ($AiFeeling)$, and interest group ($InterestGroup$). We first conducted likelihood ratio tests to test if adding each moderator improved our baseline model. Moderators were kept only if they significantly improved the fit $(p<0.05)$. See SI 7 for retained moderators. Then, we used emmeans to probe and interpret moderating effects.

\subsection{Global}
\label{global_diversity}
Intuitively, the \textit{global} diversity of ideas in a condition measures how similar or different submitted ideas in a condition tend to be. The relevant level of aggregation here is \textit{all} of the submitted ideas at a \{[\textit{condition}], \textit{item}\} level. For example, consider the total set of ideas participants submitted for a \textit{tire} in the [\textit{Control}] condition. Is this set of ideas more diverse from each other than the set of submitted ideas for a tire in the [\textit{High Exposure}, \textit{Disclosed}] condition?

We used a Monte Carlo procedure and permutation tests to assess if conditions differed with respect to these metrics. For 50 Monte Carlo runs, for each  \{[\textit{condition}], \textit{item}\} combination, we randomly sampled 50 ideas and computed idea diversity metrics. We then conducted pairwise paired (at the level of Monte Carlo seeds and items) permutation tests with 10,000 iterations to see if the two conditions differed on these metrics. As a non-parametric measure of effect size, we also calculate Cliff's Delta $(\delta)$, which ranges from -1 to 1. A value of 0 indicates no difference between the two conditions, +1 indicates values from the first condition are always \textit{larger}, and -1 indicates the opposite. See SI 8.1 for more details.

\subsection{Evolution}
\label{evo_outcomes}
\paragraph{Creativity}
To test if conditions differed in their evolution of creativity, we conducted a likelihood ratio test on whether an interaction between \textit{condition} and \textit{TrialNo} significantly improved the fit of the local creativity model. 

\paragraph{Idea Diversity}
Intuitively, we are interested in if---as the experiment goes on---ideas that participants submit tend to become more or less similar to each other. We use the trial number in a response chain to index time in the experiment. For example, is the set of submitted responses at trial number 4 more or less similar to each other as the set of submitted responses at trial number 20? Here, the diversity of interest is not between a submitted response and example responses but between all submitted responses at a given `time point' (i.e., trial number). The question is if the diversity increases or decreases as the experiment goes on and if this rate of change differs by condition. Here are the mechanics of our process. See SI Section 8.3 for more details.
\begin{enumerate}

    \item We first `pooled' together all ideas at the \{[\textit{condition}], \textit{item}, \textit{trial number}\} level, across response chains. For example, consider all ideas for a \textit{tire} for the [\textit{Control}] condition that were the \textit{fourth} response in a response chain. We refer to this set as a `pool' of ideas. 
    
    \item We next computed idea diversity measures for each pool of ideas, where idea pools were defined in (1). We use the same metrics that we measure at a local level for idea diversity. Median pairwise distance is our main measure. We conduct robustness checks using mean pairwise distance and mean distance from the centroid. Each metric shows qualitatively similar results. 

    \item We then fit a mixed model (items as random intercepts) to test if the slope of trial number on idea diversity differed by condition. That is: Are submitted responses in some conditions changing at a faster rate?

\end{enumerate}

\section{Effects of Exposure and Disclosure on Creativity and Diversity}
\label{detailed_background}
\subsection{Factor 1: LLM Exposure}

\label{exposure}
\subsubsection{Effects on Creativity}
Intuitively, the effect of exposure to ChatGPT ideas will depend on how creative ChatGPT answers are relative to human ideas. In preliminary testing, we found that the answers to the AUT generated by our prompt were scored as more creative than the ideas generated by humans (see SI 4) via the \citet{organisciak_beyond_2023} classifier. LLM generations may be increasing in creativity: while GPT-3 (an earlier model than ChatGPT-3.5) scored lower in AUT creativity than humans on the AUT \citep{stevenson_putting_2022}, GPT-4 (a more recent model than ChatGPT-3.5) scored among the \textit{top percentile} of humans on a similar verbal creativity task \cite{guzik_originality_2023}, as measured by human judges.

Even if language models can generate creative ideas, it is unclear from prior work if mere exposure to these ideas can increase human creativity. On one hand, the associative model of brainstorming suggests that exposure to others' ideas can stimulate idea generation by activating a non-accessible concept of a participant's memory \citep{nijstad_how_2006, brown_making_2002, paulus_toward_2007}. For example, ChatGPT may come up with a use for a bottle that you never associated with bottles. This can then inspire you to come up with creative uses along this line. In this way, ChatGPT can \textit{stimulate} creativity. On the other hand, there is also evidence that seeing the ideas of others \textit{inhibits} a participant's idea generation if ``one is exposed to an idea that has few connections to other ideas in an individual's semantic network'' \citep[pg. 10]{paulus_toward_2007}. Indeed, this appeared to be the case in \citet{yang_ai_2022}. There is a possibility that AI ideas are creative but so divorced from how humans generate ideas that seeing these ideas actually has an \textit{inhibiting} effect. Separate from prior \textit{academic} work, there are \textit{public} debates about the impact of LLMs (such as ChatGPT) on creativity (e.g., \citep{nation_world_news_why_2023, european_business_review_chatgpt_2023, jared_henderson_chatgpt_2022, krish_naik_will_2023,tubefilter_86_2023,eapen_how_2023,wilcot_using_2023}). Many of these debates \textit{assume} ChatGPT will have \textit{some} impact on an individual's creativity---either good or bad. Our work contributes empirical results to this broader public conversation.

\subsubsection{Effects on Diversity}
Prior work in AI co-creation finds mixed effects. Collaborating with AI can lead to more diverse \citep{yang_ai_2022, osone_buncho_2021, lee_coauthor_2022, branch_collaborative_2021, gero_metaphoria_2019} or less diverse outputs \citep{padmakumar_does_2024, doshi_generative_2024, dellacqua_navigating_2023}. But note that these studies are testing \textit{active engagement}, and most test active engagement with \textit{intentionally} constructed systems. This is different from the passive, incidental exposure to AI ideas that now occur in everyday life. Writers call ChatGPT `a blurry JPEG of the internet' \citep{chiang_chatgpt_2023} and discuss its `incredible blandness' \citep{mangalaseril_incredible_2023}; researchers call it a `stochastic parrot' \citep{bender_dangers_2021}. It is not clear, then, how passive exposure to ideas from off-the-shelf LLMs---precisely the kind we are inundated with---would affect the diversity of human ideas.

\subsection{Factor 2: LLM Disclosure}
\label{disclosure}
\subsubsection{Effects on Creativity}
Building on \citet{hwang_ideabot_2021}, we employ the theory of social facilitation \citep{bond_social_1983} to understand how LLM disclosure can affect human creativity. Facilitation theory is concerned with how the presence of others affects one's performance. \citet{hwang_ideabot_2021} asked participants to brainstorm with chatbots (which gave pre-programmed responses) and experimentally varied whether or not participants were told that their partner was a chatbot. Disclosing that the partner was a chatbot led to higher creativity in participant responses, which \citet{hwang_ideabot_2021} attributes to the novelty of brainstorming with a chatbot. We build on this notion of \textit{facilitation} as a theoretical lens.  However, it is not clear if \citet{hwang_ideabot_2021}'s finding (that telling people they are brainstorming with a chatbot increases creativity) would replicate in our study, especially in a post-ChatGPT era. First, we are measuring \textit{exposure} and not direct engagement with chatbots. The novelty of a chatbot may be higher when you are the one working with it to generate ideas. Second, presumably, the novelty of talking to a chatbot may be lower due to the widespread popularity of ChatGPT. Moreover, we may expect heterogeneity in disclosure's effect on creativity and diversity. It may be that users who have lower self-perceived creative abilities may feel `competition' with AI due to its presence and, in turn, submit more creative responses when they know the ideas they are exposed to are from AI. 

\subsubsection{Effects on Diversity}
It is not clear how knowing content is from AI will affect the diversity of ideas participants produce. But prior work suggests heterogeneity along two lines: the difficulty of the prompt\footnote{We measure the `difficulty' of a prompt by the inverse rank of the average creativity in the control condition. If participants tended to submit \textit{lower} creativity ideas in the control condition for item X, we said item X was \textit{difficult}.}, and the attitude of the participant. Prior work suggests that disclosing ideas as AI-generated would \textit{decrease} diversity due to automation bias, the tendency to over-rely on AI systems \citep{schemmer_influence_2022, mosier_automation_1996, goddard_automation_2014}. Increased reliance on AI ideas (when labeled as such) could lead to lower idea diversity and higher AI adoption. Conversely, some evidence suggests people display algorithmic aversion to creative products such as haikus \citep{hitsuwari_does_2022} or art \citep{kirk_modulation_2009}.  This aversion would yield the opposite prediction. \citet{roemmele_inspiration_2021} found that seeing AI examples only affected the participant's writing on a key measure for difficult prompts---suggesting creative task difficulty might moderate the effect of disclosure on AI adoption. Task confidence decreases reliance on automated systems and trust in a system increases reliance on automated systems \citep{goddard_automation_2014}. Although this literature is not usually applied to creativity, we might then suspect that self-reported creativity (i.e., roughly analogous to task confidence) and perceived AI creativity could affect the adoption of AI ideas when the source is disclosed.

\section{Experiment Screenshots}
\label{experiment_screenshots}
\vspace{-0.5em}
\begin{figure}[H]
   \centering
   \begin{minipage}[b]{0.45\textwidth}
       \includegraphics[width=\textwidth]{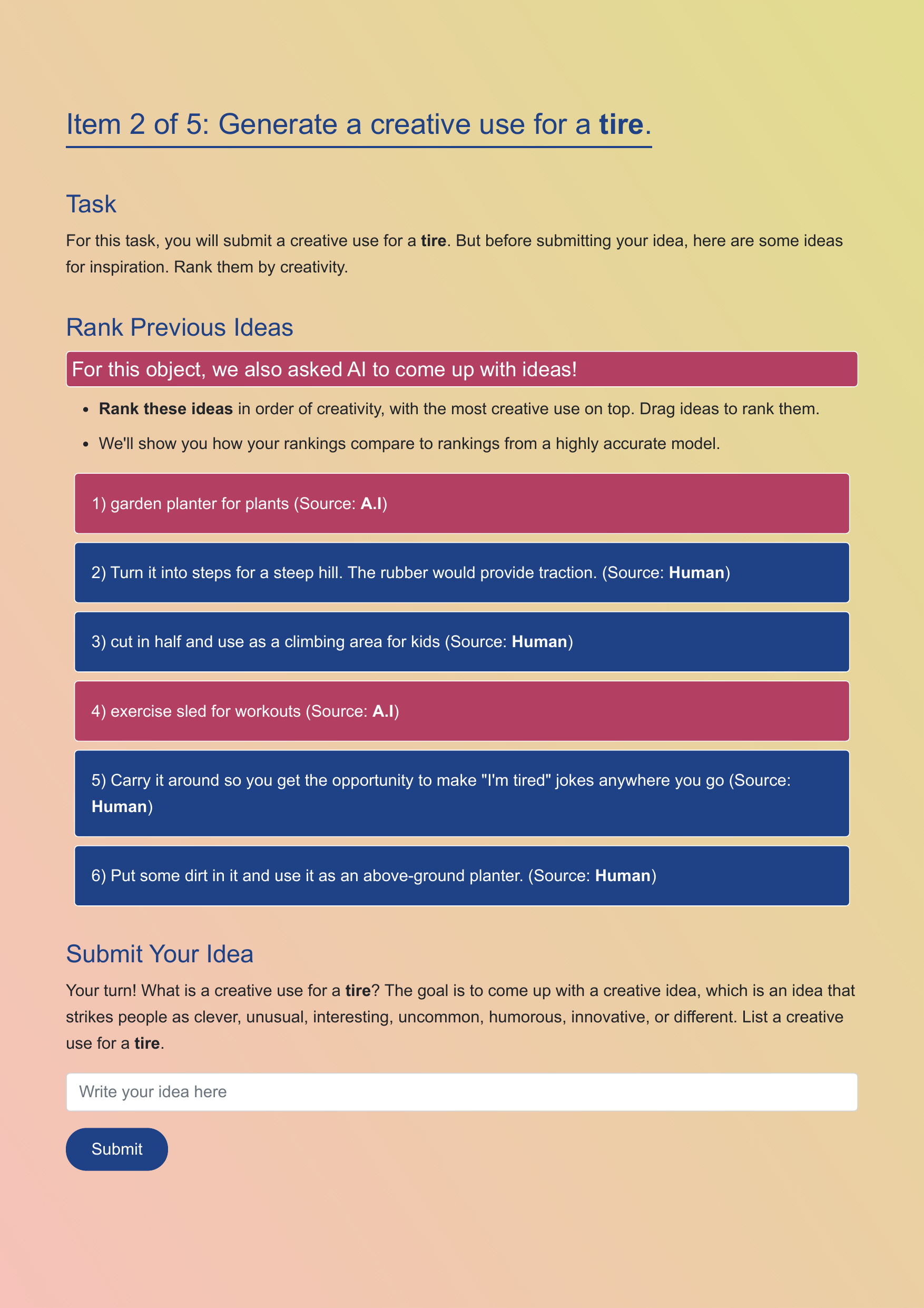}
       \caption{Screenshot of a trial participants are told to complete. This is the [\textit{Low Exposure}, \textit{Disclosed}] condition since there are two AI ideas and the AI ideas are labeled.}
       \label{fig:trial}
   \end{minipage}
   \hfill
   \begin{minipage}[b]{0.45\textwidth}
       \includegraphics[width=\textwidth]{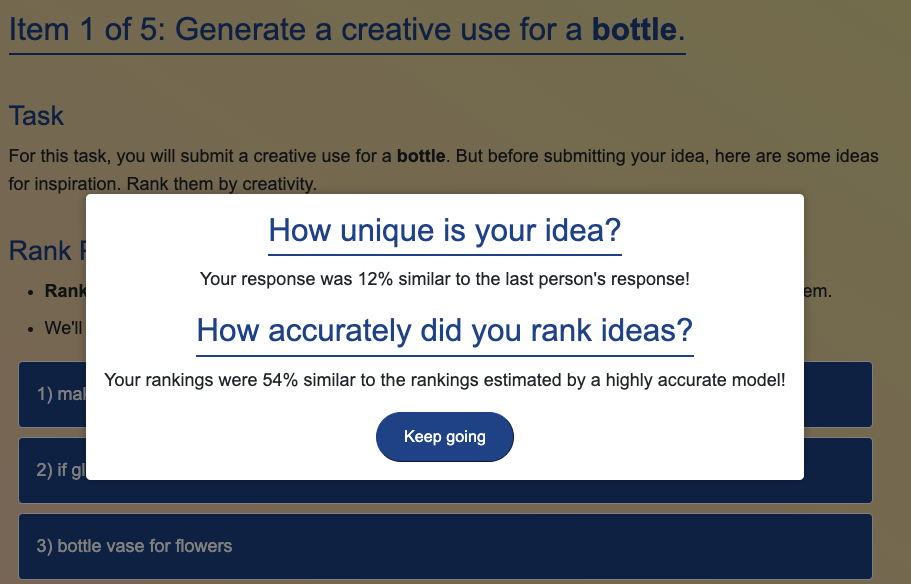}
       \caption{After each trial, participants were given feedback as motivation to continue.}
       \label{fig:feedback}
   \end{minipage}
   \label{fig:trials_and_feedback}
\end{figure}

\onecolumn
\renewcommand{\thesubsection}{SI \arabic{subsection}}
\section{Supplementary Information}

\subsection{AUT Items} 
\label{aut_items}

\begin{longtable}{lrr}
  \caption{AUT items by frequency of occurrence in dataset and classifier accuracy. Accuracy is defined as the correlation between human ratings of creativity and model predictions. The overall accuracy was r = 0.81. The accuracy of responses from the best-performing 5-item subset was r= 0.90. The data and model are from \citet{organisciak_beyond_2022}.} 
  \label{aut_desc_stats} \\
  \hline
  AUT Item & Classifier Accuracy (r) & Frequency in Test Set \\ 
  \hline
  \endfirsthead 

  \multicolumn{3}{c}{-- table continued from previous page --} \\
  \hline
  AUT Item & Classifier Accuracy (r) & Frequency in Test Set \\
  \hline
  \endhead 

  \hline \multicolumn{3}{r}{-- table continues on next page --} \\
  \endfoot 

  \hline
  \endlastfoot 

  \textbf{tire} & 0.91 & 412 \\ 
  \textbf{pants} & 0.91 & 443 \\ 
  \textbf{shoe} & 0.91 & 382 \\ 
  \textbf{table} & 0.90 & 461 \\ 
  \textbf{bottle} & 0.88 & 839 \\ 
  pencil & 0.85 & 384 \\ 
  ball & 0.84 & 393 \\ 
  fork & 0.83 & 407 \\ 
  lightbulb & 0.83 & 383 \\ 
  toothbrush & 0.81 & 379 \\ 
  knife & 0.81 & 2163 \\ 
  backpack & 0.80 &  34 \\ 
  shovel & 0.79 & 339 \\ 
  paperclip & 0.79 & 1385 \\ 
  hat & 0.76 & 380 \\ 
  box & 0.74 & 2842 \\ 
  spoon & 0.73 & 386 \\ 
  book & 0.71 & 487 \\ 
  sock & 0.69 & 380 \\ 
  brick & 0.64 & 5162 \\ 
  rope & 0.56 & 2080 \\
\end{longtable}

\begin{table}[h!]
\caption{Summary statistics of AUT prompt experiment. Human ideas are from \citet{organisciak_beyond_2022} and include only those ideas in response to the chosen AUT items. Note that in some cases, ChatGPT did not return the desired number of ideas, leading to a slight discrepancy between the ideas generated by the two prompts.}
\label{tab:aut_n_words_table}
\begin{tabular}{lrrr}
\toprule
 & N & Average Words & SD Words \\
Condition &  &  &  \\
\midrule
Human Ideas & 2537 & 4.56 & 4.97 \\
Zero Shot Length Limited & 7500 & 4.44 & 1.34 \\
Zero Shot & 8153 & 25.38 & 8.55 \\
\bottomrule
\end{tabular}
\end{table}

\subsection{Pre-Treatment Questions}
\label{pre_treatment_questions}
Pew asked about feeling towards AI \citep{nadeem_public_2023, nadeem_how_2022} and we used the specific phrasing and choice ordering from  \citep{nadeem_how_2022}. We randomized the first two options and kept neutral last. Our gender question was based on guidance from \citet{spiel_how_2019}. The options were: `woman', `man', `non-binary', `prefer to self-describe', `prefer not to disclose'. We added a text box meant for those who preferred to self-describe. The only deviation from \citet{spiel_how_2019} is that we did not allow participants to select multiple options. We note that gender (as well as age and country) was optional.

\subsection{Exclusion Criteria for Analysis}
\label{exclusion}

Participants could have consented and answered pre-treatment questions but failed to complete any trial. We only analyze data from participants who completed at least one trial. In $(n=4)$ cases, users submitted ages that were implausible. We replaced these age values with \textit{missing} for the purpose of summarizing participants but kept the responses.  In $(n=2)$ cases, responses that should not have been shown were shown. We remove these two responses from the analysis. As discussed in SI \ref{implementation_complications}, we instituted content moderation after receiving several troll responses. After the study, we manually inspected each response flagged by our system. There were $46$ ideas labeled as profane, and we determined $36$ were true positives. We remove the true positives $(n=36)$ from analysis, resulting in a final set of $3414$ responses for analysis from an initial set of $3452$ responses. Importantly, we conducted chi-squared tests and found that condition was unrelated to the number of flagged ideas ($\chi^2$(4) = 6.06, p = 0.19), number of flagged ideas minus false positives ($\chi^2$(4) = 2.92, p = 0.57) or total number of excluded ideas ($\chi^2$(4) = 3.87, p = 0.42). 
\clearpage

\subsection{Human vs AI Ideas} 
\label{human_vs_ai_ideas}
We compared a sample of 1500 ideas from our modified Stevenson prompt in the prompt experiment and a random sample of 1500 ideas from the Organisciak Dataset for our 5 items. For each set, we used the model's predicted originality scores. Originality ranges from 1-5. Overall, ChatGPT ideas had higher ($\beta = 0.62 $, $t(2994) =  22.49 $,  95\% CI = $[0.56 ,  0.67]$) originality. 

\begin{table}[!htbp] \centering 
  \caption{Comparing predicted originality of ChatGPT-generated ideas to ideas from a dataset of prior human responses} 
  \label{} 
\begin{tabular}{@{\extracolsep{5pt}}lc} 
\\[-1.8ex]\hline 
\hline \\[-1.8ex] 
 & \multicolumn{1}{c}{\textit{Dependent variable:}} \\ 
\cline{2-2} 
\\[-1.8ex] & originality \\ 
\hline \\[-1.8ex] 
 sourcechatgpt & 0.618$^{***}$ \\ 
  & (0.027) \\ 
  & \\ 
 promptpants & $-$0.072$^{*}$ \\ 
  & (0.041) \\ 
  & \\ 
 promptshoe & 0.096$^{**}$ \\ 
  & (0.043) \\ 
  & \\ 
 prompttable & $-$0.007 \\ 
  & (0.042) \\ 
  & \\ 
 prompttire & $-$0.196$^{***}$ \\ 
  & (0.041) \\ 
  & \\ 
 Constant & 2.751$^{***}$ \\ 
  & (0.029) \\ 
  & \\ 
\hline \\[-1.8ex] 
Observations & 3,000 \\ 
R$^{2}$ & 0.159 \\ 
Adjusted R$^{2}$ & 0.157 \\ 
Residual Std. Error & 0.746 (df = 2994) \\ 
F Statistic & 112.903$^{***}$ (df = 5; 2994) \\ 
\hline 
\hline \\[-1.8ex] 
\textit{Note:}  & \multicolumn{1}{r}{$^{*}$p$<$0.1; $^{**}$p$<$0.05; $^{***}$p$<$0.01} \\ 
\end{tabular} 
\end{table} 
\subsection{Participant Feedback} 
\label{participant_feedback}

We encouraged participants to start the study in the first place by saying that---if they finished all 5 trials---we would show them how creative they are relative to humans and AI. At the end of the experiment, we first computed a participant's average score from the \citet{organisciak_beyond_2022} classifier as their `creativity score'. We then graphically and verbally showed participants what percentile this score would be for both humans and AI (where the human and AI scores come from applying the \citet{organisciak_beyond_2022} classifier to a sample of AI ideas we generated and prior human ideas from the Organisciak Dataset.) We also provided a graph that compared a participant's scores in the AI condition to their scores in the no-AI conditions. 

Additionally, we wanted to minimize attrition for participants once they started. We gave participants two pieces of feedback after each trial so they would continue taking the study. 

\begin{itemize}
    \item First, we calculated how unique a participant's response was relative to the last person's response. We did this by calculating the cosine distance between a word2vec embedding of the participant's response and a word2vec embedding of the last response in a given \{[\textit{condition}], \textit{item}\}. Due to resource constraints, we used a truncated word2vec model---the top 15k words in English. 

    \item We also compared the accuracy of participants' rankings to the rankings of ideas by the classifier \citep{organisciak_beyond_2022} we used. To do this, we calculated the rank-order correlation between a participant's rankings of items and the rank order generated by the \citet{organisciak_beyond_2022} model.  
\end{itemize}
In certain cases, either of these metrics could not be calculated, and we returned an arbitrary, random number.

\subsection{Sample Characteristics}

\label{sample_characteristics}

\begin{figure}[h!]
    \centering
    \includegraphics[width=0.75\textwidth]{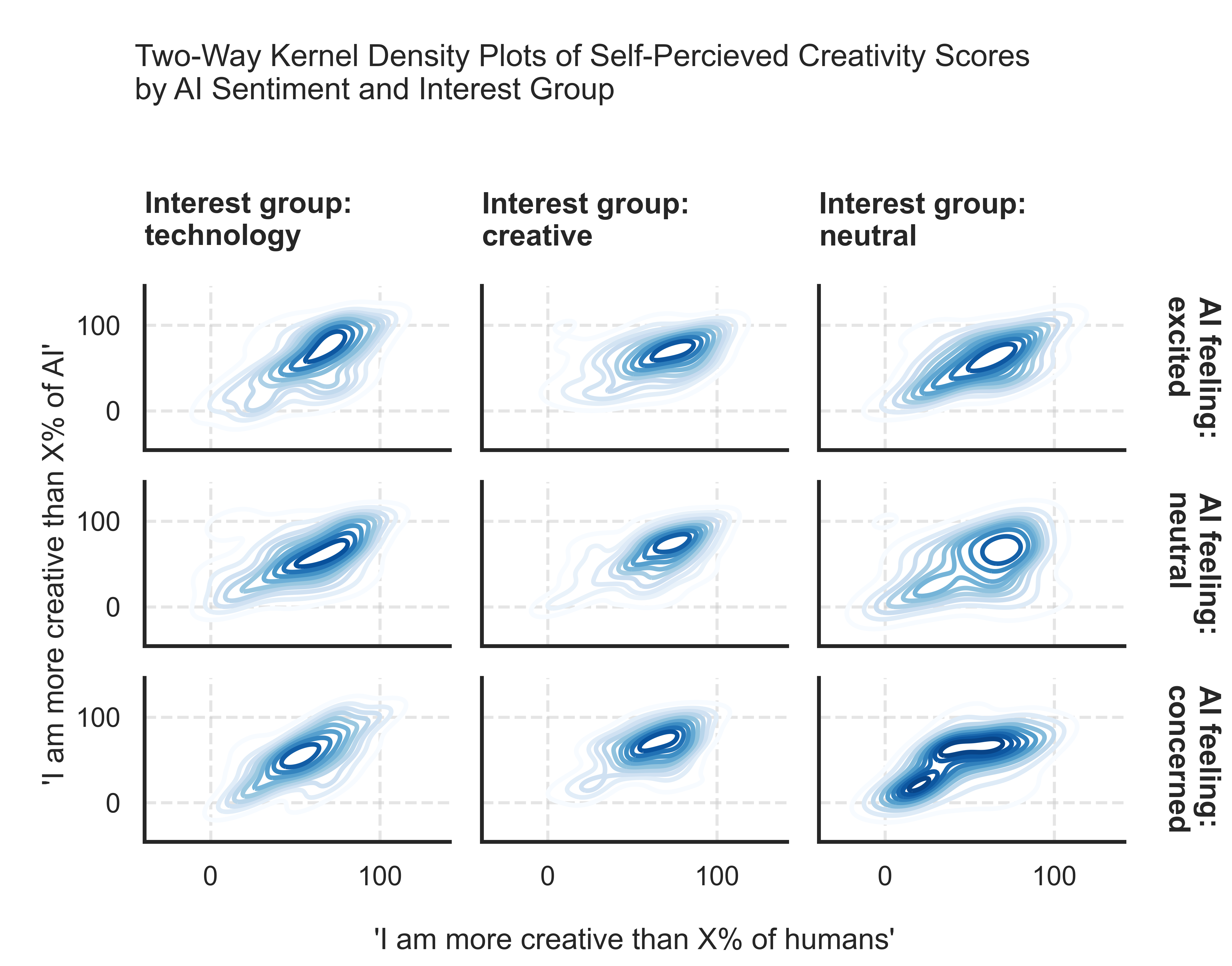}
    \caption{Distributions of self-perceived creativity relative to humans and relative to AI, by both interest group and sentiment towards AI.}
    \label{fig:two_way_hist}
\end{figure}

\begin{table}[htbp]
\centering
\caption{Descriptive Stats (Non-Missing Values)}
\label{desc_stats}
\begin{tabular}{lrrrrr}
\toprule
{} &   Mean &     SD &  25th Percentile &  Median &  75th Percentile \\
\midrule
age              &  34.92 &  10.86 &             27.0 &    33.0 &             40.0 \\
creativity\_ai    &  57.86 &  26.66 &             40.0 &    60.0 &             76.0 \\
creativity\_human &  58.67 &  23.65 &             44.0 &    62.0 &             75.0 \\
\bottomrule
\end{tabular}
\end{table}

\begin{table}[htbp]
    \centering
    \begin{minipage}{.5\textwidth}
        \centering
        \caption{Distribution of Gender}
        \label{dist_gender}
        \begin{tabular}{ll}
            \toprule
            {} & Counts (\% of total) \\
            gender               &                     \\
            \midrule
            woman                &           308 (36\%) \\
            man                  &           268 (32\%) \\
            Missing              &           222 (26\%) \\
            non-binary           &             23 (3\%) \\
            prefer\_not\_disclose  &             16 (2\%) \\
            prefer\_self\_describe &              7 (1\%) \\
            \bottomrule
        \end{tabular}
    \end{minipage}%
    \begin{minipage}{.5\textwidth}
        \centering
        \caption{Distribution of AI Feeling}
        \label{dist_ai_feeling}
        \begin{tabular}{ll}
            \toprule
            {} & Counts (\% of total) \\
            ai\_feeling &                     \\
            \midrule
            neutral    &           403 (48\%) \\
            excited    &           232 (27\%) \\
            concerned  &           198 (23\%) \\
            Missing    &             11 (1\%) \\
            \bottomrule
        \end{tabular}
    \end{minipage}
\end{table}

Although we did not assess English language proficiency, the top five countries by responses (77.85\% of responses) were the United States, Canada, Germany, United Kingdom, and Australia--- countries with high English proficiency. The median response length was six words, which is relatively short, also suggesting English language proficiency is not a likely confounder.

\subsection{Model Selection}
\label{si_model}
\begin{table}[H]
\centering
\begin{tabular}{llrrrl}
  \hline
DV & Potential Moderator & $\chi^2$ & Df & $p < \chi^2$ & Added Interaction \\ 
  \hline Idea Diversity & Self-Perceived Human Creativity & 10.32 & 4.00 & 0.04 & YES \\ 
   & AI - Human Creativity & 15.70 & 4.00 & 0.00 & YES \\ 
   & AI Feeling & 8.20 & 8.00 & 0.41 & NO \\ 
   & Interest Group & 12.02 & 8.00 & 0.15 & NO \\ 
   \hline Creativity & Self-Perceived Human Creativity & 3.28 & 4.00 & 0.51 & NO \\ 
   & AI - Human Creativity & 1.11 & 4.00 & 0.89 & NO \\ 
   & AI Feeling & 8.19 & 8.00 & 0.41 & NO \\ 
   & Interest Group & 1.57 & 8.00 & 0.99 & NO \\ 
   \hline AI Adoption & Self-Perceived Human Creativity & 18.24 & 3.00 & 0.00 & YES \\ 
   & AI - Human Creativity & 9.94 & 3.00 & 0.02 & YES \\ 
   & AI Feeling & 4.14 & 6.00 & 0.66 & NO \\ 
   & Interest Group & 13.05 & 6.00 & 0.04 & YES \\ 
  \end{tabular}
\caption{To determine which moderating variables to include, we conducted likelihood ratio tests comparing the baseline specification to a model including an interaction between a potential moderator and the treatment condition. If the likelihood ratio test indicated the interaction improved the fit at $p<0.05$, we included this interaction in our model.} 
\label{chisq_inters}
\end{table}

Selected models already include `Interest Group' to control for participant source (neutral, creative, technical). As a robustness check, we subsequently created an additional participant source variable, `IsSocialMedia', indicating if the respondent was from social media. Likelihood ratio tests found adding `IsSocialMedia' and its interaction with the treatment condition did not improve the fit of the selected models $(p>0.39 \text{ for all models})$.

\subsection{Idea Diversity}
\subsubsection{Global}
\label{si_diverge_global}
For 50 Monte Carlo runs with different seed values, we sampled 50 ideas for each \{[\textit{condition}], \textit{item}\} combination. For each 50-idea set, we computed various idea diversity measures. First, we calculated all pairwise SBERT distances. Next, we measured the mean and median pairwise distances. We also computed the centroid of each 50-idea set and calculated the mean distance from the centroid. After calculating these metrics, we conducted two-tailed, paired permutation tests (10,000 iterations) to test if two conditions differed on these metrics. To conduct the paired permutation test, we randomly swapped the sign of the difference between pairs of values, equivalent to randomly swapping the condition labels of rows---simulating the null hypothesis that conditions do not differ. We then counted the proportion of null distribution iterations where one would observe a larger absolute difference in means than the observed difference. We added a 1 to the numerator and denominator, which is a common, conservative adjustment \citep{ojala_permutation_2009} and stops p-values from being 0. Because the test is paired (equivalent to swapping the condition label within each `row'), our permutation tests are controlling for both AUT items and Monte Carlo seeds, since each row shares these attributes. We controlled for multiple pairwise comparisons by applying a Holm-Bonferroni adjustment to p-values. As a non-parametric measure of effect size, we used Cliff's Delta. This metric ranges from -1 to +1 where 0 indicates no difference between conditions, +1 indicates that all Monte Carlo runs for the first condition are larger than those for the second, and vice versa for -1. As with evolution, to avoid the confounding effect of conditions differing in the number of seeds, we consider ideas after the sixth trial (see SI \ref{si_diverge_evo} for a more detailed discussion). This is because the experiment is designed to `shed' all seed examples after trial six. 

\begin{table}[H]
\caption{Global idea diversity measured by mean pairwise distance}
\label{si_global_mean_pairwise_distance}
\begin{tabular}{llrrr}
\toprule
 & Contrast & Diff in Means & Adj P Value & Cliff's Delta \\
\midrule
0 & HighExposureDisclosed-Control & 1.340000 & 0.001000 & 0.350000 \\
1 & HighExposureUndisclosed-Control & 0.900000 & 0.001000 & 0.270000 \\
2 & LowExposureDisclosed-Control & 0.340000 & 0.011100 & 0.090000 \\
3 & LowExposureUndisclosed-Control & -0.460000 & 0.016500 & -0.020000 \\
4 & HighExposureDisclosed-HighExposureUndisclosed & 0.440000 & 0.011100 & 0.050000 \\
5 & HighExposureDisclosed-LowExposureDisclosed & 1.000000 & 0.001000 & 0.290000 \\
6 & HighExposureDisclosed-LowExposureUndisclosed & 1.800000 & 0.001000 & 0.320000 \\
7 & HighExposureUndisclosed-LowExposureDisclosed & 0.560000 & 0.001000 & 0.210000 \\
8 & HighExposureUndisclosed-LowExposureUndisclosed & 1.360000 & 0.001000 & 0.280000 \\
9 & LowExposureDisclosed-LowExposureUndisclosed & 0.800000 & 0.001000 & 0.090000 \\
\bottomrule
\end{tabular}
\end{table}

\begin{table}[H]
\caption{Global idea diversity measured by median pairwise distance}
\label{global_median_pairwise_distance}
\begin{tabular}{llrrr}
\toprule
 & Contrast & Diff in Means & Adj P Value & Cliff's Delta \\
\midrule
0 & HighExposureDisclosed-Control & 1.210000 & 0.001000 & 0.310000 \\
1 & HighExposureUndisclosed-Control & 0.810000 & 0.001000 & 0.260000 \\
2 & LowExposureDisclosed-Control & 0.470000 & 0.001000 & 0.110000 \\
3 & LowExposureUndisclosed-Control & -0.610000 & 0.006300 & -0.060000 \\
4 & HighExposureDisclosed-HighExposureUndisclosed & 0.410000 & 0.032800 & 0.040000 \\
5 & HighExposureDisclosed-LowExposureDisclosed & 0.740000 & 0.001000 & 0.230000 \\
6 & HighExposureDisclosed-LowExposureUndisclosed & 1.820000 & 0.001000 & 0.320000 \\
7 & HighExposureUndisclosed-LowExposureDisclosed & 0.330000 & 0.032800 & 0.180000 \\
8 & HighExposureUndisclosed-LowExposureUndisclosed & 1.410000 & 0.001000 & 0.290000 \\
9 & LowExposureDisclosed-LowExposureUndisclosed & 1.080000 & 0.001000 & 0.150000 \\
\bottomrule
\end{tabular}
\end{table}

\begin{table}[H]
\caption{Global idea diversity measured by mean centroid distance}
\label{global_mean_centroid_distance}
\begin{tabular}{llrrr}
\toprule
 & Contrast & Diff in Means & Adj P Value & Cliff's Delta \\
\midrule
0 & HighExposureDisclosed-Control & 1.480000 & 0.001000 & 0.350000 \\
1 & HighExposureUndisclosed-Control & 1.030000 & 0.001000 & 0.270000 \\
2 & LowExposureDisclosed-Control & 0.360000 & 0.013800 & 0.090000 \\
3 & LowExposureUndisclosed-Control & -0.430000 & 0.029500 & -0.020000 \\
4 & HighExposureDisclosed-HighExposureUndisclosed & 0.450000 & 0.013800 & 0.050000 \\
5 & HighExposureDisclosed-LowExposureDisclosed & 1.120000 & 0.001000 & 0.290000 \\
6 & HighExposureDisclosed-LowExposureUndisclosed & 1.910000 & 0.001000 & 0.320000 \\
7 & HighExposureUndisclosed-LowExposureDisclosed & 0.680000 & 0.001000 & 0.210000 \\
8 & HighExposureUndisclosed-LowExposureUndisclosed & 1.460000 & 0.001000 & 0.280000 \\
9 & LowExposureDisclosed-LowExposureUndisclosed & 0.790000 & 0.001000 & 0.090000 \\
\bottomrule
\end{tabular}
\end{table}

\clearpage
\subsubsection{Local}
\label{diversity_local_SI}
Intuitively, local idea diversity is how different a response is from the examples a participant sees. There was no main effect of condition $(F(4, 19.95) = 0.09, p = 0.98)$. We found mixed evidence that belief in AI's relative creativity moderates local idea diversity. Regression results showed a small but significant interaction effect between the [\textit{High Exposure, Undisclosed}] condition and relative AI creativity ($\beta =  0.038 $, $t(3244) =  2.149$,  95\% CI = $[0.003,0.073]$, $p=0.03$). We probed this effect with estimated marginal means, predicting local idea diversity for the bottom and top decile of participants by perception of AI creativity. Top-decile participants had slightly higher local idea diversity than bottom-decile participants in the [\textit{High Exposure, Undisclosed}] condition ($\Delta = 2.62, d = 0.34$) but although this difference was significant before multiple comparisons ($p= 0.01$), it was not significant after adjusting for multiple comparisons, ($p=0.06$; see SI Table \ref{tab:diverge_rel_c}). Hence, we conclude there is mixed evidence for the role of belief in AI's relative creativity as a moderator of local idea diversity.

\begin{table}[b] \centering 
  \caption{Predictors of local idea diversity with coefficients and SEs in parentheses. The DV for models (1) and (2) are the median and mean pairwise distances between a participant's response and examples. Model (3) uses the distance between a participant's response and the centroid of examples. Ideas are embedded using SBERT. All three models have a random intercept for participants crossed with a random intercept for response chains, nested in (item, condition) combinations.} 
  \label{reg_ai_diverge} 
\tiny 
\begin{tabular}{@{\extracolsep{5pt}}lccc} 
\\[-1.8ex]\hline 
\hline \\[-1.8ex] 
 & \multicolumn{3}{c}{\textit{Dependent variable:}} \\ 
\cline{2-4} 
\\[-1.8ex] & Median PW Distance & Mean PW Distance & Centroid Distance \\ 
\\[-1.8ex] & (1) & (2) & (3)\\ 
\hline \\[-1.8ex] 
 conditionLoExposure\_Disclosed & $-$1.467 (1.864) & $-$2.177 (1.701) & $-$3.791 (2.438) \\ 
  & t = $-$0.787 & t = $-$1.280 & t = $-$1.555 \\ 
  conditionLoExposure\_Undisclosed & 0.272 (1.861) & $-$0.098 (1.698) & $-$0.414 (2.433) \\ 
  & t = 0.146 & t = $-$0.057 & t = $-$0.170 \\ 
  conditionHiExposure\_Disclosed & 1.051 (1.864) & 0.814 (1.701) & 3.354 (2.438) \\ 
  & t = 0.564 & t = 0.479 & t = 1.376 \\ 
  conditionHiExposure\_Undisclosed & $-$0.744 (1.866) & $-$0.958 (1.703) & 0.442 (2.441) \\ 
  & t = $-$0.399 & t = $-$0.563 & t = 0.181 \\ 
  creativity\_human & $-$0.006 (0.013) & $-$0.008 (0.013) & $-$0.015 (0.021) \\ 
  & t = $-$0.484 & t = $-$0.647 & t = $-$0.723 \\ 
  ai\_rel\_create & $-$0.006 (0.013) & $-$0.005 (0.012) & $-$0.008 (0.020) \\ 
  & t = $-$0.425 & t = $-$0.401 & t = $-$0.388 \\ 
  trial\_no & $-$0.023 (0.029) & $-$0.020 (0.027) & $-$0.007 (0.045) \\ 
  & t = $-$0.814 & t = $-$0.725 & t = $-$0.166 \\ 
  ai\_feelingconcerned & 0.354 (0.358) & 0.317 (0.339) & 0.455 (0.565) \\ 
  & t = 0.990 & t = 0.935 & t = 0.807 \\ 
  ai\_feelingexcited & 0.270 (0.349) & 0.107 (0.331) & 0.207 (0.550) \\ 
  & t = 0.773 & t = 0.324 & t = 0.377 \\ 
  interest\_groupcreative & $-$0.573 (0.453) & $-$0.492 (0.431) & $-$0.694 (0.693) \\ 
  & t = $-$1.263 & t = $-$1.142 & t = $-$1.002 \\ 
  interest\_grouptechnology & 0.081 (0.465) & 0.161 (0.443) & 0.139 (0.711) \\ 
  & t = 0.173 & t = 0.363 & t = 0.196 \\ 
  condition\_order & 0.182$^{**}$ (0.092) & 0.165$^{*}$ (0.087) & 0.256$^{*}$ (0.143) \\ 
  & t = 1.982 & t = 1.897 & t = 1.798 \\ 
  log\_duration & $-$0.765$^{***}$ (0.200) & $-$0.718$^{***}$ (0.189) & $-$1.192$^{***}$ (0.313) \\ 
  & t = $-$3.820 & t = $-$3.791 & t = $-$3.809 \\ 
  n\_seeds & 0.398$^{***}$ (0.136) & 0.440$^{***}$ (0.128) & 0.597$^{***}$ (0.213) \\ 
  & t = 2.927 & t = 3.429 & t = 2.808 \\ 
  conditionLoExposure\_Disclosed:creativity\_human & 0.034$^{*}$ (0.018) & 0.040$^{**}$ (0.017) & 0.070$^{**}$ (0.028) \\ 
  & t = 1.884 & t = 2.313 & t = 2.456 \\ 
  conditionLoExposure\_Undisclosed:creativity\_human & 0.006 (0.018) & 0.005 (0.017) & 0.015 (0.028) \\ 
  & t = 0.304 & t = 0.285 & t = 0.519 \\ 
  conditionHiExposure\_Disclosed:creativity\_human & $-$0.003 (0.018) & $-$0.004 (0.017) & $-$0.013 (0.028) \\ 
  & t = $-$0.169 & t = $-$0.248 & t = $-$0.455 \\ 
  conditionHiExposure\_Undisclosed:creativity\_human & 0.024 (0.018) & 0.022 (0.017) & 0.030 (0.028) \\ 
  & t = 1.298 & t = 1.256 & t = 1.065 \\ 
  conditionLoExposure\_Disclosed:ai\_rel\_create & $-$0.018 (0.018) & $-$0.017 (0.017) & $-$0.031 (0.028) \\ 
  & t = $-$1.027 & t = $-$0.991 & t = $-$1.136 \\ 
  conditionLoExposure\_Undisclosed:ai\_rel\_create & 0.007 (0.018) & 0.005 (0.017) & 0.006 (0.028) \\ 
  & t = 0.365 & t = 0.296 & t = 0.207 \\ 
  conditionHiExposure\_Disclosed:ai\_rel\_create & $-$0.008 (0.018) & $-$0.007 (0.017) & $-$0.005 (0.027) \\ 
  & t = $-$0.462 & t = $-$0.414 & t = $-$0.199 \\ 
  conditionHiExposure\_Undisclosed:ai\_rel\_create & 0.038$^{**}$ (0.018) & 0.040$^{**}$ (0.017) & 0.065$^{**}$ (0.028) \\ 
  & t = 2.149 & t = 2.384 & t = 2.331 \\ 
  Constant & 85.104$^{***}$ (1.733) & 84.644$^{***}$ (1.607) & 72.776$^{***}$ (2.466) \\ 
  & t = 49.099 & t = 52.676 & t = 29.508 \\ 
 \hline \\[-1.8ex] 
Observations & 3,271 & 3,271 & 3,271 \\ 
Log Likelihood & $-$11,201.060 & $-$11,014.890 & $-$12,630.840 \\ 
Akaike Inf. Crit. & 22,456.120 & 22,083.790 & 25,315.680 \\ 
Bayesian Inf. Crit. & 22,620.630 & 22,248.290 & 25,480.180 \\ 
\hline 
\hline \\[-1.8ex] 
\textit{Note:}  & \multicolumn{3}{r}{$^{*}$p$<$0.1; $^{**}$p$<$0.05; $^{***}$p$<$0.01} \\ 
\end{tabular} 
\end{table}

\begin{table}
\caption{\label{tab:diverge_rel}Estimated marginal means contrasts of local idea diversity, using a mixed model to compare predictions for top 10 percentile and bottom 10 percentile of participants by belief in relative AI creativity. Local idea diversity is computed as the median pairwise distance between a participant's idea and the example ideas. P-values adjusted for multiple comparisons using the Holm-Bonferroni method.}
\centering
\resizebox{\linewidth}{!}{
\begin{tabular}[t]{lrrrrrrr}
\toprule
contrast & Relative AI Creativity Percentile & estimate & SE & df & t.ratio & Adjusted P Value & d\\
\midrule
LoExposure\_Undisclosed - HiExposure\_Undisclosed & 10 & -0.368 & 1.540 & 20.311 & -0.239 & 1.000 & -0.048\\
LoExposure\_Undisclosed - LoExposure\_Disclosed & 10 & 0.298 & 1.539 & 20.279 & 0.193 & 1.000 & 0.038\\
LoExposure\_Undisclosed - HiExposure\_Disclosed & 10 & -0.129 & 1.539 & 20.299 & -0.084 & 1.000 & -0.017\\
LoExposure\_Undisclosed - Control & 10 & 0.660 & 1.541 & 20.381 & 0.428 & 1.000 & 0.085\\
HiExposure\_Undisclosed - LoExposure\_Disclosed & 10 & 0.666 & 1.540 & 20.356 & 0.432 & 1.000 & 0.086\\
\addlinespace
HiExposure\_Undisclosed - HiExposure\_Disclosed & 10 & 0.239 & 1.539 & 20.276 & 0.155 & 1.000 & 0.031\\
HiExposure\_Undisclosed - Control & 10 & 1.029 & 1.545 & 20.585 & 0.666 & 1.000 & 0.133\\
LoExposure\_Disclosed - HiExposure\_Disclosed & 10 & -0.427 & 1.540 & 20.346 & -0.277 & 1.000 & -0.055\\
LoExposure\_Disclosed - Control & 10 & 0.363 & 1.541 & 20.390 & 0.235 & 1.000 & 0.047\\
HiExposure\_Disclosed - Control & 10 & 0.790 & 1.545 & 20.591 & 0.511 & 1.000 & 0.102\\
\addlinespace
LoExposure\_Undisclosed - HiExposure\_Undisclosed & 90 & -2.915 & 2.203 & 84.275 & -1.323 & 1.000 & -0.377\\
LoExposure\_Undisclosed - LoExposure\_Disclosed & 90 & 2.284 & 2.202 & 84.112 & 1.037 & 1.000 & 0.295\\
LoExposure\_Undisclosed - HiExposure\_Disclosed & 90 & 1.042 & 2.187 & 81.952 & 0.476 & 1.000 & 0.135\\
LoExposure\_Undisclosed - Control & 90 & 1.181 & 2.206 & 84.767 & 0.535 & 1.000 & 0.153\\
HiExposure\_Undisclosed - LoExposure\_Disclosed & 90 & 5.198 & 2.206 & 84.630 & 2.357 & 0.207 & 0.672\\
\addlinespace
HiExposure\_Undisclosed - HiExposure\_Disclosed & 90 & 3.957 & 2.188 & 81.948 & 1.809 & 0.608 & 0.512\\
HiExposure\_Undisclosed - Control & 90 & 4.096 & 2.213 & 85.683 & 1.851 & 0.608 & 0.530\\
LoExposure\_Disclosed - HiExposure\_Disclosed & 90 & -1.242 & 2.190 & 82.342 & -0.567 & 1.000 & -0.161\\
LoExposure\_Disclosed - Control & 90 & -1.102 & 2.207 & 84.791 & -0.500 & 1.000 & -0.143\\
HiExposure\_Disclosed - Control & 90 & 0.139 & 2.197 & 83.314 & 0.063 & 1.000 & 0.018\\
\bottomrule
\end{tabular}}
\end{table}
\begin{table}

\caption{\label{tab:diverge_rel_c}Estimated marginal means contrasts of local idea diversity, using a mixed model to compare predictions for the top 10 percentile and bottom ten percentile of participants by belief in relative AI creativity. Local idea diversity is computed as the median pairwise distance between a participant's idea and the example ideas. P-values adjusted for multiple comparisons using the Holm-Bonferroni method}
\centering
\resizebox{\linewidth}{!}{
\begin{tabular}[t]{llrrrrrrr}
\toprule
contrast & condition & estimate & SE & df & t.ratio & P Value & d & Adjusted P Value\\
\midrule
ai\_rel\_create10 - ai\_rel\_create90 & LoExposure\_Undisclosed & -0.077 & 1.043 & 3156.269 & -0.074 & 0.941 & -0.010 & 1.000\\
ai\_rel\_create10 - ai\_rel\_create90 & HiExposure\_Undisclosed & -2.624 & 1.040 & 3136.309 & -2.523 & 0.012 & -0.339 & 0.058\\
ai\_rel\_create10 - ai\_rel\_create90 & LoExposure\_Disclosed & 1.909 & 1.041 & 3159.288 & 1.833 & 0.067 & 0.247 & 0.268\\
ai\_rel\_create10 - ai\_rel\_create90 & HiExposure\_Disclosed & 1.094 & 1.013 & 3181.870 & 1.080 & 0.280 & 0.141 & 0.840\\
ai\_rel\_create10 - ai\_rel\_create90 & Control & 0.444 & 1.046 & 3177.974 & 0.424 & 0.671 & 0.057 & 1.000\\
\bottomrule
\end{tabular}}
\end{table}
\clearpage
\subsubsection{Evolution}
\label{si_diverge_evo}
To model the evolution of idea diversity, we pooled together submitted ideas at the level of (item, condition, trial number). We then computed the median pairwise distance, mean pairwise distance, and mean distance from centroid for each pool of ideas. We fit a model to test if idea diversity changed at a different rate for different conditions:

\begin{align*}
\text{variable}_{cti} = \beta_0 + \beta_1 \text{Condition}_{c} + \beta_2 \text{TrialNo}_{t} + \beta_3 \text{TrialNo X Condition}_{tc} + \\
&\phantom{=} \beta_4 \text{Nobs}_{cti} +u_{0i} + e_{cti}
\end{align*}

Where:
\begin{itemize}
    \item \(c\) indexes conditions.
    \item \(t\) indexes trial number.
    \item \(i\) indexes items
    \item \(\beta_0\) is the global intercept.
    \item \(u_{0i} \sim N(0, \sigma_u^2)\) are random intercepts for items
    \item \(e_{cti} \sim N(0, \sigma^2)\) is the residual
\end{itemize}

We took two additional steps to make sure our results were not driven by confounding factors. First, $Nobs$ controls for how many ideas are in the set that is being analyzed. Recall that we designed the experiment so that each (item, condition) combination was replicated exactly seven times in response chains of exactly 20 trials. However, there were some minor deviations (discussed in SI \ref{implementation_complications}) in response chains, resulting in some (item, condition, trial number) sets having fewer items than others. Hence, we control for the number of ideas in a set. Second, we only ran this analysis on data after the sixth trial in a response chain to rule out the effect of seeds on evolution. The logic here is that the condition with the most initial seeds (6) was the control condition. The experiment is designed to `shed' all seeds after trial six since by that time there would have been six experiment responses, meaning the most recent six ideas in the control condition would now all be from the experiment, and hence no seeds present in the example sets. (Note that for all \textit{local} analyses, we directly control for the number of seeds present in the example set as a fixed effect.)

\begin{table}[!htbp] \centering 
  \caption{Evolution of idea diversity by condition. Each model has a random intercept for item. The reference level for experimental conditions is the control condition.} 
  \label{sem_evo_reg} 
\tiny 
\begin{tabular}{@{\extracolsep{5pt}}lccc} 
\\[-1.8ex]\hline 
\hline \\[-1.8ex] 
 & \multicolumn{3}{c}{\textit{Dependent variable:}} \\ 
\cline{2-4} 
\\[-1.8ex] & Median PW Distance & Mean PW Distance & Centroid Distance \\ 
\\[-1.8ex] & (1) & (2) & (3)\\ 
\hline \\[-1.8ex] 
 nobs & 0.827$^{**}$ (0.337) & 0.894$^{***}$ (0.317) & 3.940$^{***}$ (0.214) \\ 
  & t = 2.454 & t = 2.818 & t = 18.450 \\ 
  conditionLow ExposureUndisclosed & $-$1.558 (3.349) & $-$1.048 (3.151) & $-$0.534 (2.121) \\ 
  & t = $-$0.465 & t = $-$0.333 & t = $-$0.252 \\ 
  conditionLow ExposureDisclosed & $-$4.087 (3.368) & $-$4.163 (3.168) & $-$1.884 (2.133) \\ 
  & t = $-$1.213 & t = $-$1.314 & t = $-$0.883 \\ 
  conditionHigh ExposureUndisclosed & $-$5.575 (3.417) & $-$4.853 (3.214) & $-$3.517 (2.164) \\ 
  & t = $-$1.632 & t = $-$1.510 & t = $-$1.625 \\ 
  conditionHigh ExposureDisclosed & $-$6.003$^{*}$ (3.416) & $-$5.573$^{*}$ (3.213) & $-$3.608$^{*}$ (2.163) \\ 
  & t = $-$1.757 & t = $-$1.734 & t = $-$1.668 \\ 
  trial\_no & $-$0.391$^{**}$ (0.175) & $-$0.321$^{*}$ (0.165) & $-$0.141 (0.111) \\ 
  & t = $-$2.232 & t = $-$1.948 & t = $-$1.273 \\ 
  conditionLow ExposureUndisclosed:trial\_no & 0.140 (0.231) & 0.111 (0.218) & 0.069 (0.146) \\ 
  & t = 0.605 & t = 0.512 & t = 0.472 \\ 
  conditionLow ExposureDisclosed:trial\_no & 0.368 (0.233) & 0.379$^{*}$ (0.220) & 0.196 (0.148) \\ 
  & t = 1.578 & t = 1.728 & t = 1.328 \\ 
  conditionHigh ExposureUndisclosed:trial\_no & 0.525$^{**}$ (0.239) & 0.461$^{**}$ (0.225) & 0.335$^{**}$ (0.151) \\ 
  & t = 2.200 & t = 2.051 & t = 2.213 \\ 
  conditionHigh ExposureDisclosed:trial\_no & 0.566$^{**}$ (0.239) & 0.561$^{**}$ (0.225) & 0.378$^{**}$ (0.151) \\ 
  & t = 2.371 & t = 2.498 & t = 2.500 \\ 
  Constant & 81.500$^{***}$ (3.899) & 78.538$^{***}$ (3.669) & 19.040$^{***}$ (2.479) \\ 
  & t = 20.904 & t = 21.404 & t = 7.681 \\ 
 \hline \\[-1.8ex] 
Observations & 362 & 362 & 362 \\ 
Log Likelihood & $-$1,158.987 & $-$1,137.571 & $-$998.930 \\ 
Akaike Inf. Crit. & 2,343.974 & 2,301.141 & 2,023.861 \\ 
Bayesian Inf. Crit. & 2,394.566 & 2,351.733 & 2,074.452 \\ 
\hline 
\hline \\[-1.8ex] 
\textit{Note:}  & \multicolumn{3}{r}{$^{*}$p$<$0.1; $^{**}$p$<$0.05; $^{***}$p$<$0.01} \\ 
\end{tabular} 
\end{table} 

\clearpage
\subsection{Creativity}
\label{si_creativity}
\begin{table}[!htbp] \centering 
  \caption{Predictors of creativity with coefficients and SEs in parentheses. This model has a random intercept for participants crossed with a random intercept for response chains, nested in (item, condition) combinations.} 
  \label{reg_creativity} 
\tiny 
\begin{tabular}{@{\extracolsep{5pt}}lc} 
\\[-1.8ex]\hline 
\hline \\[-1.8ex] 
 & \multicolumn{1}{c}{\textit{Dependent variable:}} \\ 
\cline{2-2} 
\\[-1.8ex] & Creativity \\ 
\hline \\[-1.8ex] 
 conditionLoExposure\_Disclosed & 0.011 (0.112) \\ 
  & t = 0.095 \\ 
  conditionLoExposure\_Undisclosed & 0.019 (0.112) \\ 
  & t = 0.166 \\ 
  conditionHiExposure\_Disclosed & $-$0.025 (0.113) \\ 
  & t = $-$0.221 \\ 
  conditionHiExposure\_Undisclosed & 0.050 (0.113) \\ 
  & t = 0.443 \\ 
  creativity\_human & 0.001 (0.001) \\ 
  & t = 1.587 \\ 
  ai\_rel\_create & 0.001 (0.001) \\ 
  & t = 1.506 \\ 
  interest\_groupcreative & $-$0.069$^{*}$ (0.039) \\ 
  & t = $-$1.760 \\ 
  interest\_grouptechnology & $-$0.010 (0.040) \\ 
  & t = $-$0.249 \\ 
  trial\_no & $-$0.002 (0.003) \\ 
  & t = $-$0.797 \\ 
  ai\_feelingconcerned & $-$0.017 (0.033) \\ 
  & t = $-$0.526 \\ 
  ai\_feelingexcited & $-$0.046 (0.032) \\ 
  & t = $-$1.448 \\ 
  condition\_order & 0.003 (0.008) \\ 
  & t = 0.448 \\ 
  log\_duration & 0.097$^{***}$ (0.018) \\ 
  & t = 5.525 \\ 
  n\_seeds & $-$0.014 (0.012) \\ 
  & t = $-$1.215 \\ 
  Constant & 3.192$^{***}$ (0.128) \\ 
  & t = 24.932 \\ 
 \hline \\[-1.8ex] 
Observations & 3,271 \\ 
Log Likelihood & $-$3,196.244 \\ 
Akaike Inf. Crit. & 6,430.488 \\ 
Bayesian Inf. Crit. & 6,546.252 \\ 
\hline 
\hline \\[-1.8ex] 
\textit{Note:}  & \multicolumn{1}{r}{$^{*}$p$<$0.1; $^{**}$p$<$0.05; $^{***}$p$<$0.01} \\ 
\end{tabular} 
\end{table}

\clearpage
\subsection{AI Adoption}
We measured AI adoption by the maximum cosine similarity between a participant's response and AI examples the participant saw. There was a main effect of condition $(F(3, 16.59) = 4.33, p = 0.02)$. But because we are measuring \textit{maximum} similarity, we would expect higher similarity to AI ideas in the high-exposure condition by chance (since there are more AI ideas), so we do not interpret main effects in the main manuscript and instead focus on subgroup differences.

\label{SI_adoption}
\begin{table}[ht]
\caption{\label{tab:adopt_create}Estimated marginal means contrasts of AI adoption, using a mixed model to compare predictions for top 10 percentile and bottom 10 percentile of participants by self-perceived human creativity. AI Adoption is the max cosine similarity of a participant's response and AI examples. P-values are adjusted for multiple comparisons using Holm-Bonferroni method.}
\centering
\resizebox{\linewidth}{!}{
\begin{tabular}[t]{lrrrrrrr}
\toprule
contrast & Perceived Creativity Percentile & estimate & SE & df & t.ratio & Adjusted P Value & Cohen's d\\
\midrule
LoExposure\_Undisclosed - HiExposure\_Undisclosed & 10 & -6.021 & 2.369 & 37.026 & -2.542 & 0.092 & -0.502\\
LoExposure\_Undisclosed - LoExposure\_Disclosed & 10 & -3.669 & 2.369 & 37.048 & -1.549 & 0.520 & -0.306\\
LoExposure\_Undisclosed - HiExposure\_Disclosed & 10 & -1.924 & 2.367 & 36.890 & -0.813 & 0.983 & -0.160\\
HiExposure\_Undisclosed - LoExposure\_Disclosed & 10 & 2.352 & 2.372 & 37.212 & 0.992 & 0.983 & 0.196\\
HiExposure\_Undisclosed - HiExposure\_Disclosed & 10 & 4.097 & 2.365 & 36.779 & 1.732 & 0.458 & 0.341\\
\addlinespace
LoExposure\_Disclosed - HiExposure\_Disclosed & 10 & 1.744 & 2.370 & 37.092 & 0.736 & 0.983 & 0.145\\
LoExposure\_Undisclosed - HiExposure\_Undisclosed & 90 & -5.757 & 2.172 & 26.231 & -2.650 & 0.040 & -0.480\\
LoExposure\_Undisclosed - LoExposure\_Disclosed & 90 & 0.599 & 2.170 & 26.127 & 0.276 & 1.000 & 0.050\\
LoExposure\_Undisclosed - HiExposure\_Disclosed & 90 & -6.852 & 2.168 & 26.058 & -3.160 & 0.020 & -0.571\\
HiExposure\_Undisclosed - LoExposure\_Disclosed & 90 & 6.356 & 2.174 & 26.323 & 2.924 & 0.028 & 0.529\\
\addlinespace
HiExposure\_Undisclosed - HiExposure\_Disclosed & 90 & -1.095 & 2.166 & 25.913 & -0.506 & 1.000 & -0.091\\
LoExposure\_Disclosed - HiExposure\_Disclosed & 90 & -7.451 & 2.171 & 26.171 & -3.433 & 0.012 & -0.621\\
\bottomrule
\end{tabular}}
\end{table}
\begin{table}

\caption{\label{tab:adopt_rel}Estimated marginal means contrasts of AI adoption, using a mixed model to compare predictions for the top 10 percentile and bottom 10 percentile of participants by belief in relative AI creativity. This metric captures how creative participants think AI is relative to humans (higher values mean more creative than humans). AI Adoption is the max cosine similarity of a participant's response and AI examples. P-values adjusted for multiple comparisons using the Holm-Bonferroni method.}
\centering
\resizebox{\linewidth}{!}{
\begin{tabular}[t]{lrrrrrrr}
\toprule
contrast & Relative AI Creativity Percentile & estimate & SE & df & t.ratio & Adjusted P Value & d\\
\midrule
LoExposure\_Undisclosed - HiExposure\_Undisclosed & 10 & -5.225 & 1.952 & 17.121 & -2.677 & 0.095 & -0.435\\
LoExposure\_Undisclosed - LoExposure\_Disclosed & 10 & -1.115 & 1.949 & 17.038 & -0.572 & 1.000 & -0.093\\
LoExposure\_Undisclosed - HiExposure\_Disclosed & 10 & -4.651 & 1.951 & 17.100 & -2.384 & 0.145 & -0.387\\
HiExposure\_Undisclosed - LoExposure\_Disclosed & 10 & 4.110 & 1.953 & 17.186 & 2.104 & 0.201 & 0.342\\
HiExposure\_Undisclosed - HiExposure\_Disclosed & 10 & 0.574 & 1.949 & 17.013 & 0.295 & 1.000 & 0.048\\
\addlinespace
LoExposure\_Disclosed - HiExposure\_Disclosed & 10 & -3.536 & 1.953 & 17.169 & -1.811 & 0.263 & -0.295\\
LoExposure\_Undisclosed - HiExposure\_Undisclosed & 90 & 0.422 & 3.168 & 115.486 & 0.133 & 1.000 & 0.035\\
LoExposure\_Undisclosed - LoExposure\_Disclosed & 90 & -1.388 & 3.161 & 114.586 & -0.439 & 1.000 & -0.116\\
LoExposure\_Undisclosed - HiExposure\_Disclosed & 90 & -2.363 & 3.130 & 110.316 & -0.755 & 1.000 & -0.197\\
HiExposure\_Undisclosed - LoExposure\_Disclosed & 90 & -1.811 & 3.172 & 115.993 & -0.571 & 1.000 & -0.151\\
\addlinespace
HiExposure\_Undisclosed - HiExposure\_Disclosed & 90 & -2.785 & 3.134 & 110.601 & -0.889 & 1.000 & -0.232\\
LoExposure\_Disclosed - HiExposure\_Disclosed & 90 & -0.974 & 3.135 & 110.956 & -0.311 & 1.000 & -0.081\\
\bottomrule
\end{tabular}}
\end{table}
\begin{table}
\caption{\label{tab:adopt_rel_c}Estimated marginal means contrasts of AI adoption, using a mixed model to compare predictions for top 10 percentile and bottom 10 percentile of participants by belief in relative AI creativity. This metric captures how creative participants think AI is relative to humans (higher values means more creative than humans). AI Adoption is the max cosine similarity of a participant's response and AI examples. P-values adjusted for multiple comparisons using Holm-Bonferroni method.}
\centering
\resizebox{\linewidth}{!}{
\begin{tabular}[t]{llrrrrrr}
\toprule
contrast & condition & estimate & SE & df & t.ratio & Adjusted P Value & d\\
\midrule
ai\_rel\_create10 - ai\_rel\_create90 & LoExposure\_Undisclosed & -0.987 & 1.673 & 2545.523 & -0.590 & 0.555 & -0.082\\
ai\_rel\_create10 - ai\_rel\_create90 & HiExposure\_Undisclosed & 4.660 & 1.669 & 2538.494 & 2.792 & 0.005 & 0.388\\
ai\_rel\_create10 - ai\_rel\_create90 & LoExposure\_Disclosed & -1.261 & 1.668 & 2542.774 & -0.756 & 0.450 & -0.105\\
ai\_rel\_create10 - ai\_rel\_create90 & HiExposure\_Disclosed & 1.301 & 1.611 & 2480.927 & 0.808 & 0.419 & 0.108\\
\bottomrule
\end{tabular}}
\end{table}
\begin{table}[!htbp] \centering 
  \caption{Predictors of AI adoption with coefficients and SEs in parentheses. The respective dependent variables are the max, mean, and median cosine similarities between the SBERT embedding of a participant's response and the SBERT embeddings of AI examples the participant saw. All three models have a random intercept for participants crossed with a random intercept for response chains, nested in (item, condition) combinations.} 
  \label{reg_ai_adopt} 
\tiny 
\begin{tabular}{@{\extracolsep{5pt}}lccc} 
\\[-1.8ex]\hline 
\hline \\[-1.8ex] 
 & \multicolumn{3}{c}{\textit{Dependent variable:}} \\ 
\cline{2-4} 
\\[-1.8ex] & Max AI Similarity & Mean AI Similarity & Median AI Similarity \\ 
\\[-1.8ex] & (1) & (2) & (3)\\ 
\hline \\[-1.8ex] 
 conditionLoExposure\_Undisclosed & $-$5.481$^{*}$ (2.801) & $-$3.962$^{*}$ (2.356) & $-$3.960 (2.412) \\ 
  & t = $-$1.957 & t = $-$1.682 & t = $-$1.642 \\ 
  conditionHiExposure\_Disclosed & $-$2.243 (2.799) & $-$3.498 (2.355) & $-$3.415 (2.411) \\ 
  & t = $-$0.801 & t = $-$1.485 & t = $-$1.416 \\ 
  conditionHiExposure\_Undisclosed & 3.507 (2.799) & 0.098 (2.355) & $-$0.584 (2.411) \\ 
  & t = 1.253 & t = 0.041 & t = $-$0.242 \\ 
  creativity\_human & $-$0.059$^{***}$ (0.022) & $-$0.041$^{**}$ (0.017) & $-$0.041$^{**}$ (0.017) \\ 
  & t = $-$2.726 & t = $-$2.430 & t = $-$2.363 \\ 
  ai\_rel\_create & 0.016 (0.021) & 0.015 (0.016) & 0.015 (0.017) \\ 
  & t = 0.757 & t = 0.911 & t = 0.885 \\ 
  interest\_groupcreative & 2.425$^{*}$ (1.360) & 1.853$^{*}$ (1.052) & 1.828$^{*}$ (1.074) \\ 
  & t = 1.784 & t = 1.761 & t = 1.702 \\ 
  interest\_grouptechnology & 0.708 (1.389) & 0.979 (1.074) & 0.945 (1.097) \\ 
  & t = 0.510 & t = 0.912 & t = 0.862 \\ 
  trial\_no & $-$0.037 (0.050) & 0.005 (0.039) & 0.022 (0.040) \\ 
  & t = $-$0.734 & t = 0.122 & t = 0.556 \\ 
  ai\_feelingconcerned & $-$0.261 (0.630) & $-$0.211 (0.479) & $-$0.189 (0.489) \\ 
  & t = $-$0.413 & t = $-$0.441 & t = $-$0.388 \\ 
  ai\_feelingexcited & 0.569 (0.615) & 0.165 (0.467) & 0.149 (0.477) \\ 
  & t = 0.925 & t = 0.354 & t = 0.312 \\ 
  condition\_order & $-$0.204 (0.164) & $-$0.201 (0.128) & $-$0.212 (0.131) \\ 
  & t = $-$1.247 & t = $-$1.569 & t = $-$1.619 \\ 
  log\_duration & 0.529 (0.354) & 0.539$^{**}$ (0.273) & 0.510$^{*}$ (0.279) \\ 
  & t = 1.494 & t = 1.974 & t = 1.828 \\ 
  n\_seeds & $-$0.435 (0.309) & $-$0.265 (0.240) & $-$0.214 (0.245) \\ 
  & t = $-$1.407 & t = $-$1.104 & t = $-$0.875 \\ 
  conditionLoExposure\_Undisclosed:creativity\_human & 0.053$^{*}$ (0.029) & 0.037 (0.023) & 0.037 (0.023) \\ 
  & t = 1.820 & t = 1.601 & t = 1.557 \\ 
  conditionHiExposure\_Disclosed:creativity\_human & 0.115$^{***}$ (0.029) & 0.066$^{***}$ (0.023) & 0.061$^{***}$ (0.023) \\ 
  & t = 3.931 & t = 2.868 & t = 2.601 \\ 
  conditionHiExposure\_Undisclosed:creativity\_human & 0.050$^{*}$ (0.029) & 0.030 (0.023) & 0.039$^{*}$ (0.024) \\ 
  & t = 1.702 & t = 1.293 & t = 1.655 \\ 
  conditionLoExposure\_Undisclosed:ai\_rel\_create & $-$0.003 (0.028) & $-$0.004 (0.022) & $-$0.004 (0.023) \\ 
  & t = $-$0.121 & t = $-$0.178 & t = $-$0.174 \\ 
  conditionHiExposure\_Disclosed:ai\_rel\_create & $-$0.032 (0.028) & $-$0.011 (0.022) & $-$0.007 (0.022) \\ 
  & t = $-$1.151 & t = $-$0.523 & t = $-$0.330 \\ 
  conditionHiExposure\_Undisclosed:ai\_rel\_create & $-$0.074$^{***}$ (0.028) & $-$0.065$^{***}$ (0.022) & $-$0.071$^{***}$ (0.023) \\ 
  & t = $-$2.612 & t = $-$2.924 & t = $-$3.146 \\ 
  conditionLoExposure\_Undisclosed:interest\_groupcreative & 1.036 (1.851) & 0.819 (1.446) & 0.834 (1.476) \\ 
  & t = 0.560 & t = 0.566 & t = 0.565 \\ 
  conditionHiExposure\_Disclosed:interest\_groupcreative & $-$0.465 (1.844) & $-$0.421 (1.440) & $-$0.523 (1.470) \\ 
  & t = $-$0.252 & t = $-$0.293 & t = $-$0.356 \\ 
  conditionHiExposure\_Undisclosed:interest\_groupcreative & $-$3.224$^{*}$ (1.843) & $-$2.329 (1.440) & $-$2.344 (1.470) \\ 
  & t = $-$1.750 & t = $-$1.617 & t = $-$1.595 \\ 
  conditionLoExposure\_Undisclosed:interest\_grouptechnology & 2.810 (1.888) & 1.516 (1.474) & 1.535 (1.504) \\ 
  & t = 1.488 & t = 1.028 & t = 1.020 \\ 
  conditionHiExposure\_Disclosed:interest\_grouptechnology & $-$1.391 (1.871) & $-$1.550 (1.461) & $-$1.731 (1.491) \\ 
  & t = $-$0.743 & t = $-$1.061 & t = $-$1.161 \\ 
  conditionHiExposure\_Undisclosed:interest\_grouptechnology & $-$1.522 (1.875) & $-$1.870 (1.465) & $-$1.926 (1.495) \\ 
  & t = $-$0.811 & t = $-$1.277 & t = $-$1.288 \\ 
  Constant & 23.789$^{***}$ (2.704) & 17.655$^{***}$ (2.185) & 17.609$^{***}$ (2.235) \\ 
  & t = 8.796 & t = 8.080 & t = 7.880 \\ 
 \hline \\[-1.8ex] 
Observations & 2,618 & 2,618 & 2,618 \\ 
Log Likelihood & $-$10,155.720 & $-$9,508.078 & $-$9,561.834 \\ 
Akaike Inf. Crit. & 20,371.430 & 19,076.160 & 19,183.670 \\ 
Bayesian Inf. Crit. & 20,547.540 & 19,252.260 & 19,359.770 \\ 
\hline 
\hline \\[-1.8ex] 
\textit{Note:}  & \multicolumn{3}{r}{$^{*}$p$<$0.1; $^{**}$p$<$0.05; $^{***}$p$<$0.01} \\ 
\end{tabular} 
\end{table}

\clearpage
\subsection{Implementation Complications} \label{implementation_complications}
Running a massive online networked experiment open to any user on the Internet will often invite implementation challenges. In the interest of disclosure---and for the benefit of future researchers running similar experiments---we share the challenges we faced, solutions we implemented, and rationales for our decisions.  

\subsubsection{Server Capacity}
Overall, we received far more responses than we expected. At several points throughout the experiment, we experienced more concurrent traffic than the application was designed to handle. Hence, we had to temporarily turn off the experiment to wait out high demand, add more resources, or implement and test changes described in Content Moderation. We note that these capacity issues did not affect the responses we collected. 

\subsubsection{Content Moderation}
Initially, we did not implement any content moderation. But on July 8th, we received an influx of responses from Reddit. Several participants were trolls, providing repetitive, profane responses. We then implemented a form of content moderation, flagging any idea that contained a word in a list of words banned by Google as of July 8, 2023\footnote{https://github.com/coffee-and-fun/google-profanity-words} and subsequently added two more words to the list. If an idea was flagged, it was written to our database but not shown to future participants. Some profane ideas were already shown to participants in between the time when the responses were submitted and we saw and implemented our solution. Also, the content moderation strategy was imperfect: 10 of 46 flagged ideas were false positives. As discussed earlier, condition was unrelated to the number of flagged ideas ($\chi^2$(4) = 6.06, p = 0.19) or total number of excluded ideas ($\chi^2$(4) = 3.87, p = 0.42).

We acknowledge there are more advanced and nuanced content moderation strategies, but this one was the best option given our specific circumstances and constraints. First, this bag-of-words method is very transparent. Second, we deployed this experiment through Heroku, which imposes a CPU limit on the project, precluding the use of pre-trained classifiers such as BERT. Third, we did not want to use APIs like Jigsaw or OpenAI moderation endpoints because these APIs have rate limits, which can slow down the experiment. 

\subsubsection{Small Deviations From 20 Trials Per Chain}

We intended for each response chain to contain 20 responses. The average number of trials per response chain was 19.73 (SD = 1.45) and the median number per chain was 20. We concluded the experiment before the last round of response chains was completely finished for all condition and item combinations, so the minimum number of trials in a response chain (occurring for an item and condition combination in the last round) was 14. The maximum number of trials was 24. These minor deviations occurred due to server overload, very high traffic leading to race conditions and excluding several responses based on the criteria described in \ref{exclusion}. Based on a two-way ANOVA, we concluded that response chain lengths did not differ by item, $(F(4) = 0.38, p = 0.82)$, condition $(F(4) = 0.01, p = 1.00)$, or the interaction between items and conditions $(F(16) = 0.01, p = 1.00)$,

\end{document}